\documentclass[twocolumn,floatfix,superscriptaddress]{revtex4-2} 
\usepackage[utf8]{inputenc}
\usepackage{graphicx}
\usepackage{bigstrut}
\usepackage{amsmath}
\usepackage{xcolor}
\usepackage{multirow}
\usepackage{hyperref}
\hypersetup{colorlinks=true,allcolors=blue}
\usepackage{comment}
\usepackage{orcidlink}% Inserting a linked ORCiD logo beside author's name
\usepackage{enumitem}
\usepackage{placeins}
\usepackage[super]{nth}

\newcommand{\rev}[1]{\textcolor{black}{#1}}

\begin{document}

\title{Efficient first principles based modeling via machine learning: from simple representations to high entropy materials}

\author{Kangming Li\orcidlink{0000-0003-4471-8527}}
\email[]{kangming.li@utoronto.ca}
\affiliation{
Department of Materials Science and Engineering, University of Toronto, 27 King’s College Cir, Toronto, ON, Canada.
}

\author{Kamal Choudhary\orcidlink{0000-0001-9737-8074}}
\affiliation{Material Measurement Laboratory, National Institute of Standards and Technology, 100 Bureau Dr, Gaithersburg, MD, USA.}

\author{Brian DeCost\orcidlink{0000-0002-3459-5888}}
\affiliation{Material Measurement Laboratory, National Institute of Standards and Technology, 100 Bureau Dr, Gaithersburg, MD, USA.}

\author{Michael Greenwood}
\affiliation{Canmet MATERIALS, Natural Resources Canada, 183 Longwood Road south, Hamilton, ON, Canada.}

\author{Jason Hattrick-Simpers\orcidlink{0000-0003-2937-3188}}
\email[]{jason.hattrick.simpers@utoronto.ca}
\affiliation{
Department of Materials Science and Engineering, University of Toronto, 27 King’s College Cir, Toronto, ON, Canada.
}
\affiliation{Acceleration Consortium, University of Toronto. 80 St George St, Toronto, ON M5S 3H6.}
\affiliation{Vector Institute for Artificial Intelligence, 661 University Ave, Toronto, ON, Canada.}
\affiliation{Schwartz Reisman Institute for Technology and Society, 101 College St, Toronto, ON, Canada.}

\begin{abstract}
High-entropy materials (HEMs) have recently emerged as a significant category of materials, offering highly tunable properties. However, the scarcity of HEM data in existing density functional theory (DFT) databases, primarily due to computational expense, hinders the development of effective modeling strategies for computational materials discovery. In this study, we introduce an open DFT dataset of alloys and employ machine learning (ML) methods to investigate the material representations needed for HEM modeling. Utilizing high-throughput DFT calculations, we generate a comprehensive dataset of 84k structures, encompassing both ordered and disordered alloys across a spectrum of up to seven components and the entire compositional range. We apply descriptor-based models and graph neural networks to assess how material information is captured across diverse chemical-structural representations. We first evaluate the in-distribution performance of ML models to confirm their predictive accuracy. Subsequently, we demonstrate the capability of ML models to generalize between ordered and disordered structures, between low-order and high-order alloys, and between equimolar and non-equimolar compositions. Our findings suggest that ML models can generalize from cost-effective calculations of simpler systems to more complex scenarios. Additionally, we discuss the influence of dataset size and reveal that the information loss associated with the use of unrelaxed structures could significantly degrade the generalization performance. Overall, this research sheds light on several critical aspects of HEM modeling and offers insights for data-driven atomistic modeling of HEMs.
\end{abstract}

\maketitle

\section{Introduction}
The quest for novel materials is pivotal in advancing technology and addressing global challenges. High-entropy materials (HEMs), characterized by their multiple principal elements and high chemical disorder, have emerged as an important material class. Their remarkable properties have attracted significant attention, leading to applications in diverse areas such as catalysis, batteries, and hydrogen storage~\cite{yeh2018breakthrough, miracle2017critical, george2019high, ma2021high, amiri2021recent, sun2021high}.

The vast design space of HEMs offers immense opportunities but also poses substantial challenges in material discovery. Data-driven approaches, particularly machine learning (ML), have been increasingly employed to explore this space~\cite{rickman2020machine, qiao2021focused, chen2022high, li2020high}. While numerous studies have focused on developing ML models using experimental data, these datasets are relatively small, typically ranging from hundreds to thousands of data points, and cover only a limited portion of the potential design space~\cite{huang2019machine, kaufmann2020searching, zhou2023machine}.

High-throughput density functional theory (DFT) calculations have become a key method for generating extensive materials data. Recent efforts have led to the curation of several large DFT datasets encompassing millions of materials~\cite{curtarolo2012aflow, jain2013commentary, Saal2013, choudhary2020joint, Chanussot2021}. However, these datasets lack representations for HEMs due to the prohibitive computational cost to simulate chemical disorder~\cite{ikeda2019ab}. The computational expense of DFT methods typically scales with $\mathcal{O}(N^3)$, where $N$ is the system size. While ordered phases can be represented by small primitive cells, disordered phases require theoretically infinite primitive cell due to the aperiodic atomic arrangement. Structures with over thousands of atoms randomly distributed on the lattice may be considered sufficiently representative of HEMs, but this approach is computationally intensive and rarely used. The standard approach is the use of special quasirandom structures (SQSs) to approximate the correlation function of disordered phases, allowing the simulation of disordered materials with down to tens of atoms. Yet, the computational cost of SQSs is still significantly higher than that for ordered structures~\cite{zunger1990special, van2013efficient, yang2023high, li2020ground}.

The inclusion of SQSs in the next phase of high-throughput DFT databases development has been proposed~\cite{shen2022reflections}. However, the prioritization of specific SQS data types and their implications remain unresolved questions. Most existing studies have focused on equimolar compositions~\cite{bokas2021unveiling,chen2023map,sarker2018high,kaufmann2020discovery,lederer2018search,jiang2016efficient}, while non-equimolar compositions, which could offer superior material performance, are largely underexplored~\cite{sorkin2022first}. ML models trained on equimolar data have been applied to non-equimolar compositions without further DFT validation~\cite{vazquez2022efficient,zhang2022design}. The generalization performance of these ML models may be severely degraded~\cite{li2022critical,li2023exploiting}, due to the highly biased sampling of local chemical environments in a single equimolar material. In addition, it is unclear whether data of binary compounds are sufficient to extrapolate to HEMs with five or more elements~\cite{chen2023map}, or if the use of multi-element SQSs is necessary for systematic accuracy improvement. Furthermore, there is an accuracy-efficiency trade-off regarding the size of SQSs. A larger SQS can better approximate the disordered phase but requires substantially more compute. The suitable SQS size for the HEM modeling is closely related to the composition range and number of elements under consideration. Deviating from equimolar concentration or increasing the number of alloying elements would require larger SQSs to mimic the statistics of a random structure~\cite{jiang2016efficient,sorkin2022first}. It is therefore important to discuss the choices of the SQS parameters such as composition, number of elements, and system size in the HEM modeling.

Traditionally, ordered structures and SQS data are often exclusively used in the HEM modeling~\cite{sarker2018high,kaufmann2020discovery,jiang2016efficient,sorkin2022first,lederer2018search,bokas2021unveiling,chen2023map,vazquez2022efficient,zhang2022design}. One reason is that these methods are specifically designed to use only either ordered structures or SQSs and thus are inflexible in terms of the choice of structure sets. \rev{In contrast, many ML methods provide general representations of crystal structures independent of chemical order, which offer greater flexibility and potentially higher accuracy and efficiency~\cite{schutt2014represent,faber2015crystal,isayev2017universal,ward2017including,ward2018matminer,ong2013python,choudhary2022recent}.} However, there is a lack of ML studies that treat ordered structures and SQSs on equal footing.

\rev{
In this work, we focus on formation energy as it is crucial for thermodynamic stability assessment. The DFT-experimental deviation of formation energy is estimated to be 0.05 to 0.25 eV/atom~\cite{kingsbury2022flexible,stevanovic2012correcting,bartel2019role,kirklin2015open}, comparable to the experimental variability of 0.08 eV/atom~\cite{kirklin2015open}. While formation energy prediction has been a main target for ML studies (more commonly on DFT~\cite{bartel2022review,griesemer2023accelerating,gong2022calibrating} than on experimental data~\cite{gong2022calibrating,mao2021prediction}), it is largely limited to ordered structures due to the lack of SQS data~\cite{bartel2022review,griesemer2023accelerating,gong2022calibrating}.
}

\rev{
To address these gaps, we perform high-throughput DFT calculations to curate a large dataset for HEMs. Our SQS-containing dataset includes approximately 84k alloys with 2 to 7 components, spanning a wide range of system sizes, chemical orders, and compositions}. We examine the effects of various factors on the predictive accuracy of descriptor-based ML models and graph neural networks. Particularly, we focus on their out-of-distribution generalization capabilities, i.e., the ability to generalize to structures with different characteristics (such as more complex structures) than the training data. We first assess the predictive capabilities of ML models based on in-distribution performance. Then, we evaluate their out-of-distribution generalization capabilities, including comparisons between ordered structures and SQSs, low-order and high-order systems, and equimolar and non-equimolar compositions. Additionally, we discuss the impact of training data size and quantify information loss associated with using unrelaxed structures.

\section{Computational details}
\subsection{DFT calculations}
DFT calculations were conducted using the Vienna Ab-initio Simulation Package (VASP) code~\cite{Kresse1993a,Kresse1996e,Kresse1996c}. We used the Perdew-Burke-Ernzerhof (PBE) generalized gradient approximation for the exchange-correlation functional~\cite{Perdew1996a} and a plane-wave basis cutoff of 520 eV. We adopted the Methfessel-Paxton broadening scheme with a smearing width of 0.1 eV~\cite{Methfessel1989}. The electronic convergence cutoff was set to $10^{\text{-}5}$ eV/atom. Structures were fully relaxed to an energy convergence criterion of $10^{\text{-}4}$ eV/atom. For $k$-point sampling, grids with a density exceeding 1000 $k$-points per reciprocal atom were utilized following the Monkhorst-Pack scheme~\cite{Monkhorst1976a}. All calculations were spin-polarized. The Pymatgen package was used for input file generation and data analysis~\cite{ong2013python}. Body-centered cubic (bcc) and face-centered cubic (fcc) structures were generated using the Alloy Theoretic Automated Toolkit~\cite{van2002alloy}: ordered structures (2 to 8 atoms) through structure enumeration, and disordered structures (27, 64, or 125 atoms) using the SQS approach~\cite{van2013efficient}. Formation energies were calculated relative to the most stable unary phases of the constituent elements.

\subsection{ML modeling\label{sec:ml}}
Here we consider XGBoost (XGB)~\cite{xgboost}, random forest (RF)~\cite{breiman2001random,sklearn}, and the Atomistic LIne Graph Neural Network (ALIGNN)~\cite{choudhary2021atomistic}. XGB and RF are tree ensembles that use compositional and structural descriptors extracted from atomic structures based on the Voronoi tessellation featurization scheme~\cite{ward2017including} implemented in the Matminer package~\cite{ward2018matminer}. We use a descriptor set that consists of 145 compositional features~\cite{Ward2016} related to stoichiometry, element properties, valence orbital shells, and ionic properties, and 128 structural features~\cite{ward2017including} including statistics of coordination numbers, chemical ordering, local difference of element properties, and variance in the bond lengths and atomic volumes. ALIGNN is a graph neural network that explicitly encodes bond angle information~\cite{choudhary2021atomistic}. These models are representative of the state-of-the-art performance of descriptor-based models and graph neural networks based on their performance in various prediction tasks in the JARVIS learderboard~\cite{choudhary2023large}.

We used the following hyperparameters for XGB, RF and ALIGNN models. \rev{These hyperparameters have been used in our previous work showing consistently good model performance across various materials datasets~\cite{li2023exploiting,choudhary2023large}, and they were also found to be suitable for our dataset based on the hyperparameter grid search (Supplemental Material).} For the RF model, we disabled bootstrapping, used 100 estimators, 30~\% of the features for the best splitting, and default settings (\verb|scikit-learn| version 1.3.0~\cite{sklearn}) for other hyperparameters. For the XGB model, we used a forest with 6 parallel trees, 500 boosting rounds, a learning rate of 0.4, an L1 and L2 regularization strength of 0.01 and 0.1 respectively, the histogram tree grow method, a subsample ratio of columns to 0.5 when constructing each tree, and a subsample ratio of columns to 0.7 for each level. For the ALIGNN model, we used 2 ALIGNN layers, 2 GCN layers, a batch size of 32, and layer normalization, while keeping other hyperparameters the same as in the original ALIGNN implementation~\cite{choudhary2021atomistic}. We trained the ALIGNN model for 50 epochs as we found additional training provided negligible performance improvement. We used the OneCycle learning rate scheduler~\cite{smith2018superconvergence}, with 30~\% of the training budget allocated to linear warmup and 70~\% to cosine annealing. 

The ML models were trained with either relaxed structures or unrelaxed structures as input structures, and the formation energies of the relaxed structures as the prediction target. While the relaxed structures were obtained from DFT structural relaxations and can have local and global lattice distortions, the unrelaxed structures have an ideal bcc or fcc lattice with each atom sitting exactly on the lattice site. In addition, the lattice parameters of unrelaxed structures were set to be the same for ML training. Therefore, ML models trained with unrelaxed structures only learn from the atomic configuration in a way similar to on-lattice models~\cite{sanchez1984generalized,lederer2018search,wang2020combining,li2022magnetochemical}. 

\section{Results and discussion}

\subsection{DFT formation energy dataset}
Our DFT dataset encompasses bcc and fcc structures composed of Cr, Mn, Fe, Co, Ni, Cu, Al, and Si. The $3d$ transition-metal elements are chosen because they are the main components of Cantor alloys, one of the most important family of HEMs, with the bcc and fcc phases being the most relevant disordered phases. Al and Si are included due to their potential in enhancing corrosion resistance and developing durable materials for clean energy infrastructure~\cite{hattrick2023designing}. The dataset covers all possible 2- to 7-component alloy systems formed by these eight elements. For each alloy system, the dataset covers ordered structures and SQSs over the entire concentration range. The concentration step is 1/$n$ for ordered structures with $n$ ($2\leq n \leq 8$) atoms, and is equal to 11.1~\%, 12.5~\%, and 8~\% for SQSs with 27, 64, and 125 atoms, respectively. Table~\ref{tab:nelements} gives an overview of the numbers of alloy systems and structures. \rev{More details on the dataset can be found in the Supplemental Material.}

\begin{table}
\caption{Numbers of alloy systems and structures. We refer to structures with 2 to 8 atoms as ordered structures, and structures with 27, 64, or 125 atoms as SQSs.}
\label{tab:nelements}
\begin{ruledtabular}
\begin{tabular}{llllllll}
No. components    & 2    & 3     & 4     & 5     & 6    & 7    & Total \\
\hline
Alloy systems & 28   & 56    & 70    & 56    & 28   & 8    & 246   \\
Ordered           & 4975 & 22098 & 29494 & 6157  & 3132 & 3719 & 69575 \\
SQS               & 715  & 3302  & 3542  & 4718  & 1183 & 762  & 14222 \\
Ordered+SQS    & 5690 & 25400 & 33036 & 10875 & 4315 & 4481 & 83797 \\
\end{tabular}
\end{ruledtabular}
\end{table}

\begin{figure}
    \centering
    \includegraphics[width=\linewidth]{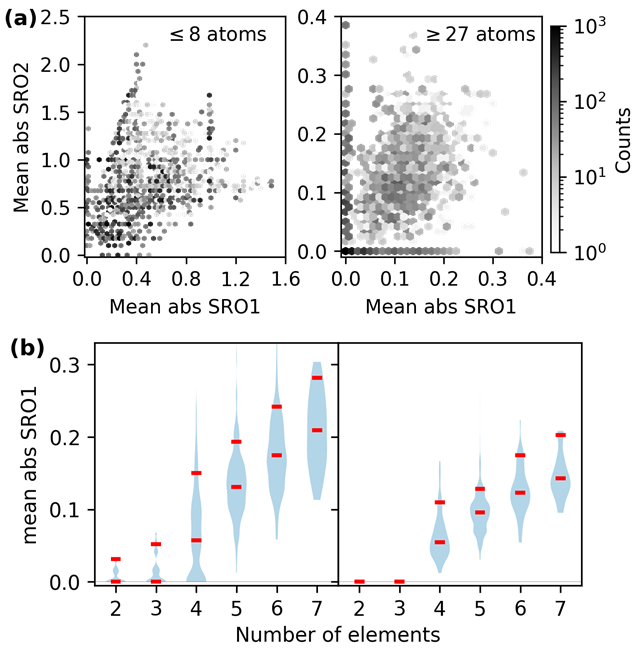}%{violinplot_SRO1_vs_nelements.png}
    \caption{
    (a) Distribution of ordered structures and SQSs in the chemical ordering space. The $X$ and $Y$ axis are the absolute SRO parameters averaged over all the chemical pairs in the first and second coordination shells. (b) Distribution of SRO parameters as a function of number of elements in 27-atom (left panel) and 64-atom (right panel) SQSs. The horizontal marks indicate the \nth{50} and \nth{90} percentiles.}
    \label{fig:violinplot_SRO1_vs_nelements}
\end{figure}

The combination of ordered structures and SQSs can enable diverse sampling of chemical order, which is quantified by the Warren-Cowley short-range order (SRO) parameter for a pair of species $i$ and $j$~\cite{Cowley1950}: 
\begin{align}
    \alpha_{ij}(r) = 1 - \frac{c_j^i(r)}{c_j} = 1 - \frac{c_i^j(r)}{c_i} = \alpha_{ji}(r)
\end{align}
where $c_j$ is the nominal concentration of the species $j$, and $c_j^i(r)$ is the concentration of the species $j$ in the $r$-th shell around the species $i$. The magnitude of $\alpha_{ij}(r)$ quantifies the degree of random mixing between the two species, with $\alpha_{ij}(r)=0$ indicating an ideal mixing. 
Here we quantify the overall chemical ordering for the $r$-th shell by averaging the magnitudes of SRO parameters for all  pairs of distinct species $i$ and $j$ in an $n$-component alloy:
\begin{align}
    \alpha(r) = \frac{ \sum_{i=1}^{n} \sum_{j=i+1}^{n} |\alpha_{ij}(r)| }{ \frac{n(n-1)}{2} }
\end{align}
In Fig.~\ref{fig:violinplot_SRO1_vs_nelements}(a), we show the distribution of overall chemical ordering for the first two shells. Ordered structures span a wide range of the SRO space but are poorly represented in the region with low SRO that is characteristic of the chemical disorder of HEMs.
The low SRO portion of our dataset predominantly consists of SQS data. %Fig.~\ref{fig:violinplot_SRO1_vs_nelements}(a) therefore illustrates the limitation of ordered structures and the advantage of SQS data in sampling the chemical ordering space of HEMs. 

Some of the generated SQSs in Fig.~\ref{fig:violinplot_SRO1_vs_nelements}(a) have non-zero SRO ranging between 0.1 to 0.3. Although SQSs with non-zero SRO are beneficial for modeling SRO effects in HEMs, it is worth emphasizing that the original intent of the SQS method was to minimize SRO to enable unbiased studies of chemically disordered systems. The significant SRO in this SQS dataset reflects the challenges in fully capturing chemical disorder within the SQS framework. As depicted in Fig.\ref{fig:violinplot_SRO1_vs_nelements}(b), the minimal SRO parameters increases with the number of elements for a given system size. For a specific system size and element count (e.g., 27-atom SQSs with two elements), higher SRO is associated with SQSs deviating further from equimolar concentration. This is because increasing the number of components reduces the number of atoms per element (deviating from concentrated compositions has a similar effect on the minority elements), hence making it difficult to arrange atoms to create diverse local environments in an SQS. Although using 64-atom SQSs can reduce the average SRO of binary and ternary structures to 0, further reducing the SRO in structures with more components or stronger non-equimolar deviation would require even larger system sizes. This highlights the computational challenge of directly simulating complex HEMs within the DFT-SQS framework, especially considering the $\mathcal{O}(N^3)$ scaling of DFT costs.

\subsection{Generalization performance with relaxed structures}

\begin{figure*}
    \centering
    \includegraphics[width=0.95\linewidth]{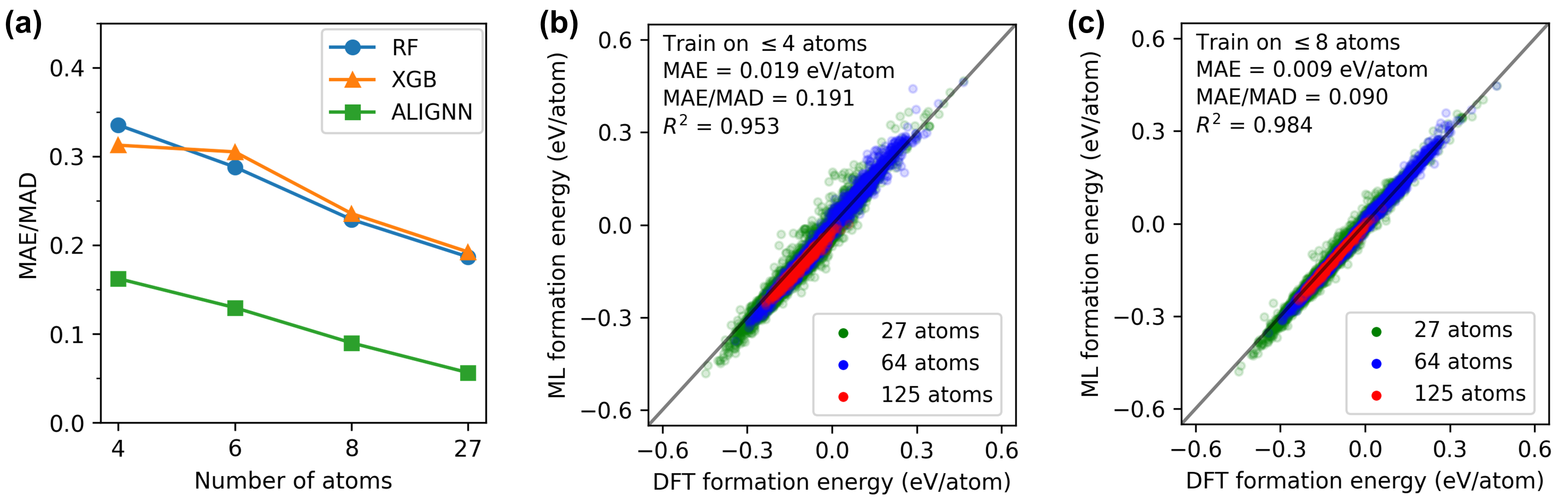}
    \caption{Generalization performance from small to large structures. (a) Normalized error obtained by training on structures with $\leq N$ atoms and evaluating on structures with $>N$ atoms. (b) Parity plot of the ALIGNN prediction on SQSs with $\geq27$ atoms, obtained by training on structures with $\leq 4$ atoms. (c) Parity plot of the ALIGNN prediction on SQSs with $\geq27$ atoms, obtained by training on structures with $\leq 8$ atoms.
    }
    \label{fig:small2large}
\end{figure*}

ML models can be used as surrogates of DFT calculations to efficiently screen the vast HEM space. To demonstrate their predictive ability, we train and evaluate the model performance with a random 8:2 train-test split of the whole dataset. We used mean absolute error (MAE) and normalized error (MAE normalized by the mean absolute deviation, MAD, of test data) as performance metrics. MAD quantifies the statistical fluctuation of the test data and can be seen as the MAE of a baseline model that always predicts the mean of the test data. A model with a normalized error below 0.2 is often considered a good predictive model~\cite{choudhary2021atomistic,gong2022examining,Dunn2020}. The RF and XGB models achieve an MAE of 0.016 eV/atom and 0.014 eV/atom, respectively, and a normalized error of 0.147 and 0.128, respectively, demonstrating the good predictive ability of the tree-based models. Compared to the tree-based models, the ALIGNN model achieves an even lower MAE of 0.007 eV/atom, or a normalized error of 0.064. \rev{The better ALIGNN performance is attributed to the use of deep graph neural networks for automated feature extraction and the explicit incorporation of bond angle information, consistent with the superior performance of graph neural networks over shallow ML methods seen in benchmark studies~\cite{Dunn2020,choudhary2021atomistic}. }

The training and test sets created via random splitting are expected to follow the same statistical distribution, and the examined performance is referred to as the in-distribution performance. In practical applications, however, ML models often encounter new data that do not necessarily follow the same distribution as the training data. It is therefore crucial to evaluate whether trained ML models can generalize beyond their original data distribution. Such an assessment not only provides an estimation of the extended applicability domain of the models but also sheds light on the types of data that should be prioritized in future DFT calculations for improved predictions.

Here we evaluate the out-of-distribution performance by training on a specific group of structures and evaluating on the rest of structures. The grouping criteria considered here are based on the system size, the number of elements, and the composition. Given that most literature DFT data comprise small-sized structures with few elements and/or equimolar composition, we focused on whether models trained on these simpler materials could generalize to more complex ones.

We first assessed the models' ability to generalize predictions from small to large structures. Fig.~\ref{fig:small2large}(a) presents normalized errors for models trained on structures with $\leq N$ atoms and tested on structures with $>N$ atoms. Remarkably, the ALIGNN model trained on structures with $\leq 4$ atoms exhibited good performance ($\frac{MAE}{MAD}= 0.16$) on structures with $> 4$ atoms. This is notable considering the limited chemical orders and compositions in the training set compared to the diverse chemical space of the 5- to 125-atom structures in the test set. Including larger structures in the training set systematically improved ALIGNN performance, reducing its normalized error on large SQSs to 0.05 when trained on structures with $\leq$ 27 atoms. RF and XGB models also benefited from including larger structures, though their normalized errors remained significantly higher than that of the ALIGNN model.

Fig.~\ref{fig:small2large}(b) shows the distribution of ALIGNN prediction errors on SQSs. The ALIGNN model trained on structures with $\le$ 4 atoms achieves a MAE of 0.019 eV/atom on SQSs and the performance is consistently good for SQSs of different sizes. The ALIGNN performance on SQSs can be further improved with a 53~\% decrease in the MAE when ordered structures with up to 8 atoms are added into the training set, as shown in Fig.~\ref{fig:small2large}(c). This good generalization performance suggests that existing large DFT databases containing mainly ordered structures could be a good starting point for HEM modeling.

\begin{figure*}
    \centering
    \includegraphics[width=0.95\linewidth]{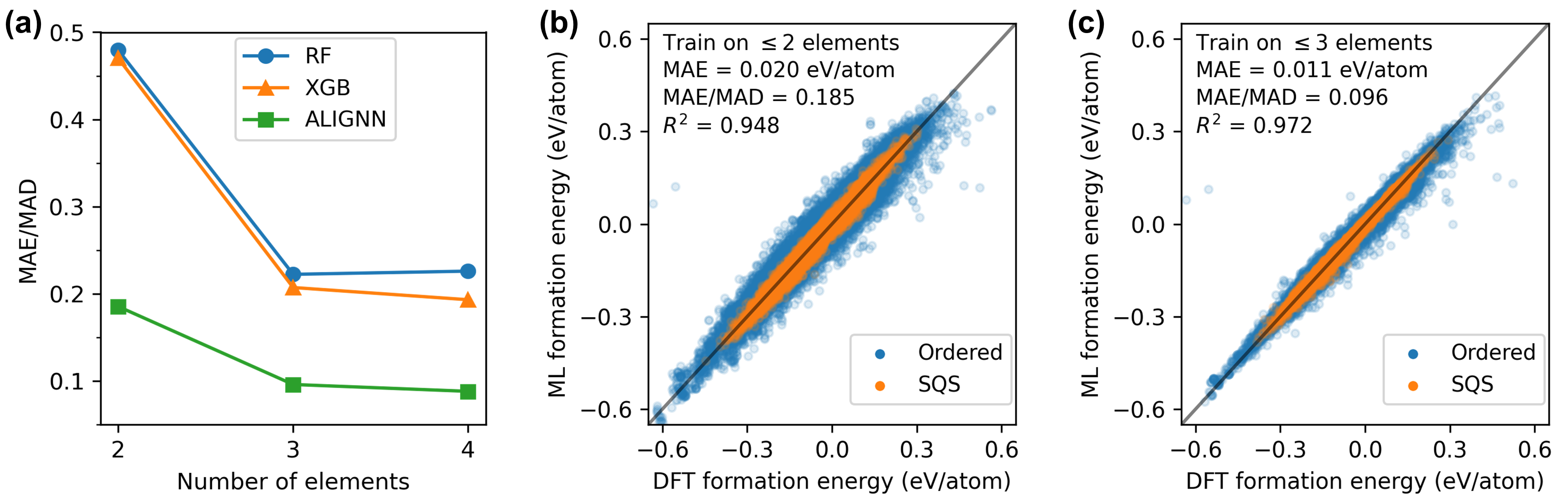}
    \caption{Generalization performance from low-order to high-order structures. (a) Normalized error obtained by training on structures with $\leq N$ elements and evaluating on structures with $>N$ elements. (b) Parity plot of the ALIGNN prediction on structures with $\geq 3$ elements, obtained by training on binary structures. (c) Parity plot of the ALIGNN prediction on structures with $\geq 4$ elements, obtained by training on binary and ternary structures.
    }
    \label{fig:nele}
\end{figure*}

Next, we examine the generalization performance from low-order to high-order systems. This is motivated by the availability of lower-order systems in current DFT datasets and the difficulty to rely on DFT to explore the vast high-order materials space. As shown in Fig~\ref{fig:nele}(a), the ALIGNN model trained on binary alloys can generalize reasonably well to alloys with three or more components, with a normalized error below 0.2 which is much lower than those obtained with the RF and XGB models. Additionally including ternary alloys into the training data can significantly improve the model performance than training only binary structures. However, this enhancement in accuracy reaches a plateau when structures with four or more components are included in the training dataset, indicating diminishing returns in accuracy boost from expanding the complexity of the training structures. Fig.~\ref{fig:nele}(b) and (c) show the parity plots for the ALIGNN predictions on structures with more than 2 components and more than 3 components, respectively. 
\rev{
In both cases, the prediction errors tend to be larger for ordered structures than for SQSs, which may be related to the fact that ordered structures can present very distinct and diverse chemical order compared to SQSs (Fig.~\ref{fig:violinplot_SRO1_vs_nelements}). 
These results suggest that SQSs may be an easier target than ordered structures for ML models in terms of achieving good accuracy, in contrast to the fact that SQSs are traditionally considered as more complex representations and also computationally heavier for DFT calculations.
}

It is worth noting that the good generalization from low-order to high-order systems is not surprising. Indeed, from a physical perspective, low-order systems are expected to contain sufficient information of many-body interactions to describe high-order systems. For instance, it is common practice in the Calphad community to develop multicomponent databases based on the thermodynamic assessment of binary and ternary systems~\cite{zhang2016calphad,chen2018database}. 
Therefore, contrary to the recent claim that this is an emergent out-of-distribution generalization enabled by advanced neural network architectures~\cite{merchant2023scaling}, we argue that this capability to generalize from low-order to high-order systems should be common to various ML models, including the traditional tree ensembles shown here. 

\begin{figure*}
    \centering
    \includegraphics[width=0.95\linewidth]{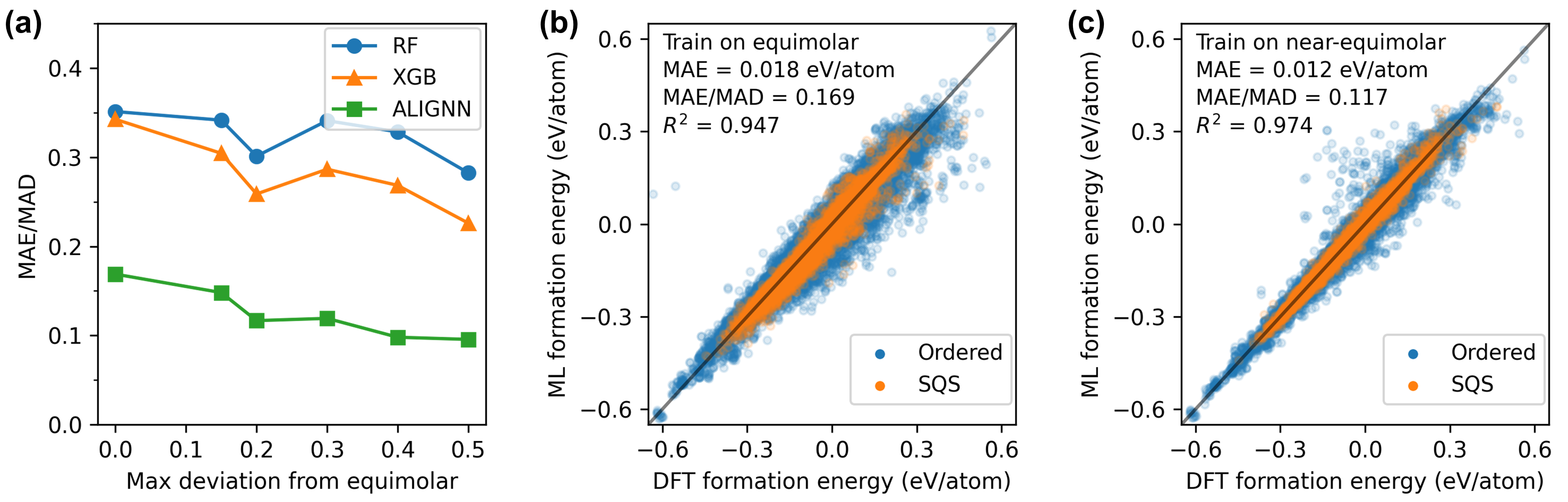}
    \caption{
    Generalization performance from (near-)equimolar to non-equimolar structures. (a) Normalized error obtained by training on structures with $\max\Delta c$ below a given threshold and evaluating on the rest.  
    Predictions on non-equimolar structures ($\max \Delta c > 0$) by the ALIGNN model trained on equimolar structures ($\max \Delta c = 0$). (c-d) Predictions on structures with relatively strong deviation from equimolar composition ($\max \Delta c > 0.2$) by the ALIGNN model trained on structures with relatively weak deviation from equimolar composition ($\max \Delta c \le 0.2$). $\max \Delta c$ is defined as the maximum concentration difference between any two elements in a structure. 
    }
    \label{fig:max_diff_c_0}
\end{figure*}

The third type of the out-of-distribution performance is based on alloy compositions. Studies focusing on HEMs have predominantly concentrated on equimolar compositions~\cite{bokas2021unveiling,chen2023map,sarker2018high,kaufmann2020discovery,lederer2018search,jiang2016efficient}, leaving a comparative dearth of data on non-equimolar compositions. Here we explore the extent to which models trained on equimolar alloys could extend their predictive capabilities to non-equimolar counterparts. Additionally, we systematically discern the impact of varying the concentration range covered in the training data. To this end, we quantify the deviation from the equimolar composition by $\max\Delta c$, defined as the maximum concentration fraction difference between any two elements in a structure. Setting a maximum value for $\max\Delta c$ is equivalent to setting the concentration range. For instance, with $\max\Delta c\le0.2$ (20~\%), the atomic fractions of all elements would fall within the 10~\% range from the equimolar composition. 

We evaluate the model performance by training on all structures with $\max\Delta c$ below a given threshold value and testing on remaining structures. The results for the threshold value of 0 are shown in Fig.~\ref{fig:max_diff_c_0}(a-b). The ALIGNN model trained on only equimolar alloys achieves a good performance with a normalized error of 0.169 for non-equimolar alloys. It is worth noting that the number of equimolar structures (7.7k data) only accounts for 9~\% of the whole dataset. Moreover, only 2~\% of the equimolar data structures SQSs while the rest of the training data are ordered structures. Despite the constrained quantity and compositional range of the training data, the ALIGNN model exhibits a robust capability to generalize to non-equimolar alloys, in particular for SQSs. Expanding the training set to include near-equimolar structures are found to improve the model performance. As demonstrated in Fig.~\ref{fig:max_diff_c_0}(c), setting the threshold to 0.2, wherein the training dataset represents 40~\% of the entire dataset, resulted in a normalized error of 0.117, or a 30~\% reduction in error compared to the equimolar case. However, further incrementing the threshold value beyond 0.2 yielded marginal improvements in performance, suggesting a saturation point in the efficacy of expanding the concentration range within the training data.

\subsection{Effects of data size and use of unrelaxed structures on model performance}

\begin{figure*}
    \centering
    \includegraphics[width=\linewidth]{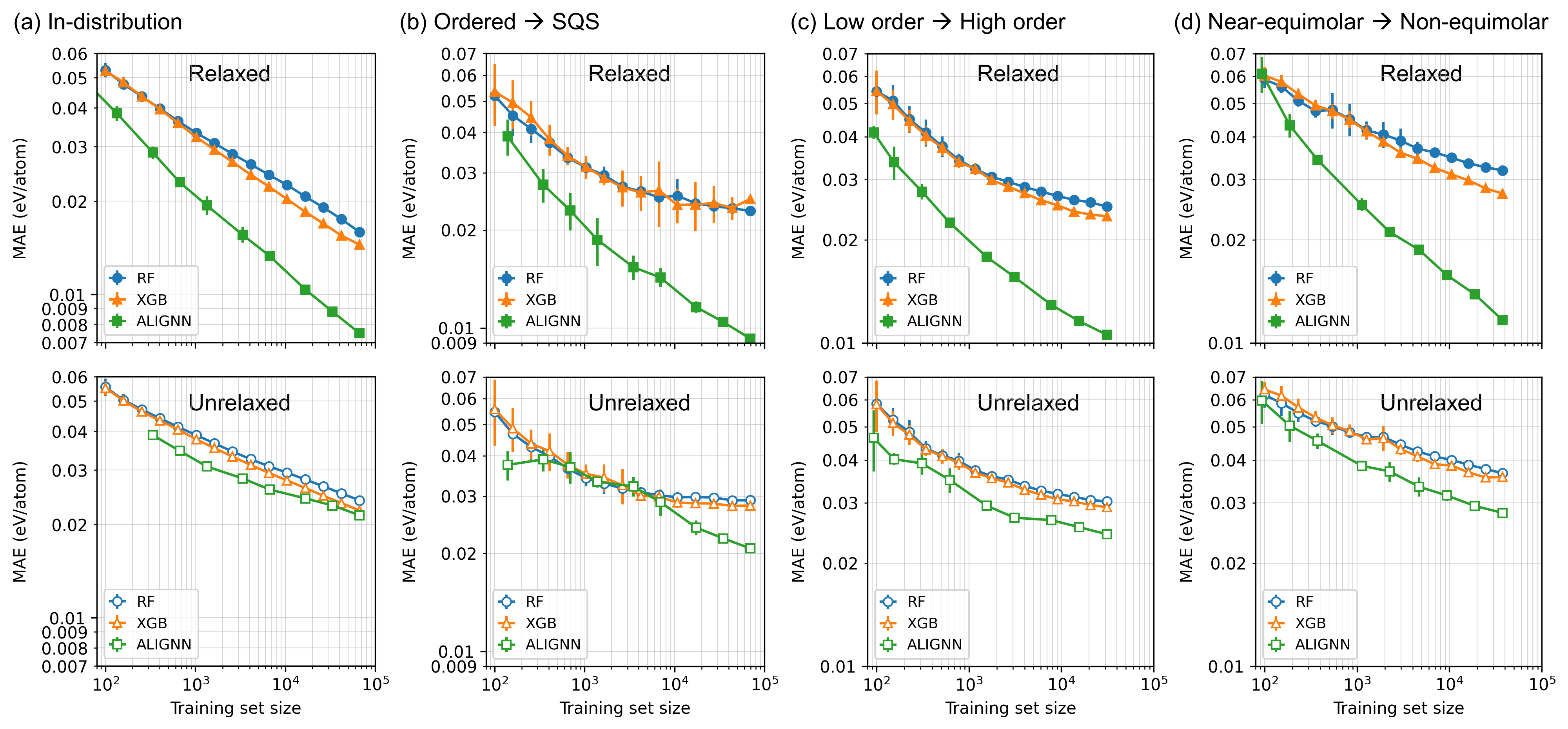}
    \caption{
    Effects of training set size and structural relaxation on the model performance. (a) In-distribution performance obtained from the random train-test splitting. (b) Performance obtained by training on ordered structures with $\leq$ 8 atoms and testing on SQSs with $\geq$ 27 atoms. (c) Performance obtained by training on binary and ternary structures and testing on structures with $\geq$ 4 components. (d) Performance obtained by training on near-equimolar structures with $\max \Delta c \leq 0.2$ and testing on other non-equimolar structures with $\max \Delta c > 0.2$. The upper and lower panels are the results obtained with relaxed and unrelaxed structures, respectively.
    }
    \label{fig:size_effect}
\end{figure*}
Fig.~\ref{fig:size_effect} reveals the effects of training set size on the model performance. In both in-distribution and out-of-distribution generalization tasks, the ALIGNN model exhibits superior performance across both small-data and large-data regimes when trained on relaxed structures. Notably, ALIGNN demonstrates a more favorable performance-versus-data scaling compared to the tree ensembles, with its performance advantage amplifying alongside the augmentation of the training set size. A compelling example of this is depicted in Fig.\ref{fig:size_effect}(b), where ALIGNN models trained on merely about 600 ordered structures reach the same MAE on SQSs as tree ensembles trained on a dataset over $100\times$ larger (69.6k ordered structures). This contrasts with the performance saturation of tree ensembles observed beyond 7k ordered structures, while ALIGNN's MAE persistently decreases with an increased number of training data.

When trained on the unrelaxed structures, all the ML models show the degraded performance compared to the ones obtained with the relaxed structures. The ALIGNN model, in particular, undergoes the most pronounced degradation, with its performance advantage narrowing to a margin only slightly better than that of the tree ensembles. In addition, the performance-versus-data scaling of the ALIGNN model also becomes similar to that of the tree ensembles. 

The performance degradation here is likely due to the information loss in the training data: when trained on unrelaxed structures, ML models are confined to leveraging solely on-lattice configurational information for energy mapping, thereby losing access to information such as local structural distortions and variations in cell shape and volume. Therefore, in scenarios where only unrelaxed structures are accessible, tree ensembles might already capture the majority of the learnable information from the training data. Consequently, graph neural networks like ALIGNN may not necessarily yield substantial performance enhancements over descriptor-based models in these settings. This underscores that the distinct advantage of graph neural networks is considerably diminished without the availability of relaxed structures for training. Conversely, the results achieved with relaxed structures underscore the significant amount of learnable information embedded in structural relaxation and the enhanced capability of graph neural networks to effectively utilize this information when available. This highlights the critical role of structural details in model training and the potential of advanced neural network models to extract deeper insights from complex structural data in materials science.

\subsection{Comparison between models across generalization tasks}

The previous sections are focused on the generalizations from ordered to disordered structures, from low-order to high-order alloys, and from equimolar compositions to non-equimolar compositions. Equally pertinent, however, is the exploration of these generalizations in reverse: from disordered to ordered structures, from high-order to low-order alloys, and from non-equimolar to equimolar compositions. Such reverse generalizations are particularly relevant in a top-down modeling approach, where one might e.g. start with data on complex high-order HEMs and aim to extrapolate to simpler, lower-order systems.

Table~\ref{tab:OOD performance} shows a comparison of the model performance for the in-distribution and six out-of-distribution tasks. When trained on relaxed structures, ALIGNN consistently outperforms the tree ensembles across all generalization tasks, achieving up to a 60~\% reduction in MAE. However, this performance disparity between ALIGNN and the tree-based models becomes less pronounced when training is based on unrelaxed structures.

\begin{table*}
\caption{
Performance for different generalization tasks and ML models. The ID column gives the in-distribution MAE, and the columns $A\rightarrow B$ give the out-of-distribution generalization MAE with $A$ as the training set and $B$ as the test set. The column labels have the following meanings. Ordered: ordered structures with $\leq$ 8 atoms, SQS: SQSs with $\ge 27$ atoms, Low: low-order structures with $\leq$ 3 elements, High: high-order structures with $\ge$ 4 elements, Equi: equimolar structures ($\max \Delta c = 0$), Non-equi: non-equimolar structures ($\max \Delta c > 0$). The second last row indicates the MAD of the test set. The last row indicates the number of training data.
}
\label{tab:OOD performance}
\begin{ruledtabular}
\begin{tabular}{ccccccccc}
Structure  & Model & ID & Ordered$\rightarrow$SQS & SQS$\rightarrow$Ordered & Low$\rightarrow$High & High$\rightarrow$Low & Equi$\rightarrow$Non-equi &Non-equi$\rightarrow$Equi \\
\hline
\multirow{3}{*}{Relaxed}     & RF     &0.016 & 0.023       & 0.040       & 0.025    & 0.038 & 0.037 & 0.029 \\
                             & XGB    &0.014 & 0.025       & 0.038       & 0.023    & 0.036 & 0.036 & 0.025 \\
                             & ALIGNN &0.007 & 0.009       & 0.025       & 0.011    & 0.013 & 0.018 & 0.010 \\ 
                             \hline
\multirow{3}{*}{Unrelaxed}   & RF     &0.024 & 0.029       & 0.043       & 0.030    & 0.043  & 0.043 & 0.036 \\
                             & XGB    &0.022 & 0.028       & 0.042       & 0.029    & 0.041  & 0.041 & 0.032 \\
                             & ALIGNN &0.021 & 0.021       & 0.043       & 0.024    & 0.030  & 0.032 & 0.026 \\ 
                             \hline
\multicolumn{2}{c}{MAD}                         & 0.109 &0.101 &0.104 & 0.111 & 0.097 & 0.106 & 0.130 \\
\multicolumn{2}{c}{Training set size}& 67k& 70k & 14k & 6k & 78k & 8k & 76k
% $67.0k\rightarrow16.7k$ & $69.6k\rightarrow14.2k$ & $14.2k\rightarrow69.6k$ & $5.7k\rightarrow78.1k$ & $78.1k\rightarrow5.7k$ & $7.7k\rightarrow76.1k$ & $76.1k\rightarrow7.7k$
\end{tabular}
\end{ruledtabular}
\end{table*}

Compared to the generalization from ordered to disordered structures, the generalization from disordered to ordered structures is more difficult, evidenced by up to a 180~\% increase in MAE. The better generalization in the former case could be largely due to the data diversity for ordered structures (Fig.~\ref{fig:violinplot_SRO1_vs_nelements}a), rather than the larger training set size of ordered structures, considering that the performance scaling with data size is relatively modest (as shown in Fig\ref{fig:size_effect}b). 

Compared to the generalization from equimolar to non-equimolar structures, the generalization from non-equimolar to equimolar is easier with a lower MAE. This outcome aligns with expectations, given the significantly larger dataset and the greater diversity found in non-equimolar compositions.

Compared to the generalization from low-order to high-order alloys, the generalization from high-order to low-order alloys is more difficult. However, this cannot be simply explained by the dataset size nor the data diversity. Indeed, the number of high-order structures is more than an order of magnitude larger than that of low-order structures. Furthermore, structures with four or more elements are expected to provide more diverse sampling of chemical environments than  binary and ternary structures. Instead, this difficulty might be explained from a physics perspective. When building physics-based model Hamiltonians, a common practice is to use a bottom-up approach, where one first fits the two-body interaction parameters with low-order systems, and gradually adds more model parameters by including higher-order interactions for more complex systems~\cite{li2022magnetochemical}. This is often considered to be physically more reasonable than just performing a single fit for all the model parameters once. The latter can be prone to overfitting by inappropriately attributing more contribution to the higher-order interactions, whereas the bottom-up approach provides a way to add regularization to higher-order interactions. In a similar spirit, the ML models trained on low-order structures can learn well low-order interactions and thus generalize to high-order systems even without knowing high-order interactions. By contrast, training only on high-order systems means the models need to decode simultaneously the low-order and high-order interactions from high-order structures, which could make the learning more difficult and would require more data than the bottom-up approach.

\section{Discussion}

\rev{
The out-of-distribution generalization results can provide insights for future DFT dataset construction. For instance, we demonstrate that ordered structures contain sufficient information for ML models to generalize well to SQSs, highlighting the usefulness of the existing DFT databases as a good starting point for HEM modeling. Furthermore, we reveal that continuously adding more complex representation would not be an efficient strategy to systematically improve the generalization performance. For example, further including quaternary systems into the binary and ternary training data only improves marginally model accuracy. These results call for the design of effective sampling strategies that take into account both the usefulness and the cost of a data point. Namely, one should weigh between an extensive but indirect sampling with many inexpensive calculations of ordered structures and a direct sampling of SQSs. There is a similar trade-off between an extensive sampling of many low-order systems and a direct sampling of a few high-order systems.
}

\rev{
We also reveal that the unavailability of DFT-relaxed structures cause significant performance degradation. This degradation is attributed to the loss of learnable information related to lattice distortion in the unrelaxed data rather than the intrinsic limitation of ML models. With unrelaxed structures, graph neural networks do not have a significant performance advantage over tree ensembles, which may be explained by the reduced amount of learnable information being a limiting factor. One possible solution is to develop ML interatomic potentials by training on the relaxation trajectory data and then use the trained ML interatomic potentials to relax the structures. However, this would incur higher training cost and also additional compute cost to perform additional simulations for relaxations. On the other hand, ML models trained on unrelaxed structures can be seen as on-lattice models~\cite{sanchez1984generalized,lederer2018search,wang2020combining,li2022magnetochemical}, which are widely used for thermodynamic and kinetic modeling thanks to their high efficiency with respect to off-lattice models or interatomic potentials. Therefore, from the cost-effective perspective, one interesting line of research may be to develop descriptor-based on-lattice models, since their accuracy is similar to neural networks, to study thermodynamics such as chemical order and phase diagram for HEMs.
}

\section{Conclusion}
\rev{
In summary, we create a DFT dataset of formation energies for 84k alloys with up to 7 components and a diverse range of concentrations and chemical orders. We find a good in-distribution performance of ML models on the dataset, with a best MAE of 0.007 eV/atom. Furthermore, we systematically investigate the generalizability of ML models between different types of structures, revealing that models trained on simpler alloy systems can generalize well to more complex ones. In addition, we analyze the effects of dataset size, and highlight the performance degradation due to the unavailability of relaxed structures. We believe these results, with our publicly available datasets and ML models, can provide valuable insights for the first principles based modeling of HEMs. 
}
\section*{Data availability}
The curated DFT dataset is publicly available on Zenodo at \url{https://doi.org/10.5281/zenodo.10854500}.

\section*{Code availability}
The code used for ML training and analysis is publicly available on GitHub at \url{https://github.com/mathsphy/high-entropy-alloys-dataset-ML}.

\section*{Conflicts of interest}
There are no conflicts of interest to declare.

\section*{Acknowledgments}
\rev{
This research was undertaken thanks in part to funding provided to the University of Toronto's Acceleration Consortium from the Canada First Research Excellence Fund (Grant number: CFREF-2022-00042).} 
The computations were made on the resources provided by the Calcul Quebec, Westgrid, and Compute Ontario consortia in the Digital Research Alliance of Canada (alliancecan.ca), and the Acceleration Consortium (acceleration.utoronto.ca) at the University of Toronto. 
We acknowledge partial funding provided by Natural Resources Canada’s Office of Energy Research and Development (OERD). 
Certain commercial products or company names are identified here to describe our study adequately. Such identification is not intended to imply recommendation or endorsement by the National Institute of Standards and Technology, nor is it intended to imply that the products or names identified are necessarily the best available for the purpose.

\section*{Author contributions}
K.L. conceived the project, performed the DFT calculations, dataset curation, ML training and analysis, and drafted the manuscript. J.H.-S. supervised the project. K.L., K.C, B.D, M.G, and J.H.-S. discussed the results. All authors reviewed and edited the manuscript, and contributed to the manuscript preparation.

\bibliography{lib}

\begin{thebibliography}{74}%
\makeatletter
\providecommand \@ifxundefined [1]{%
 \@ifx{#1\undefined}
}%
\providecommand \@ifnum [1]{%
 \ifnum #1\expandafter \@firstoftwo
 \else \expandafter \@secondoftwo
 \fi
}%
\providecommand \@ifx [1]{%
 \ifx #1\expandafter \@firstoftwo
 \else \expandafter \@secondoftwo
 \fi
}%
\providecommand \natexlab [1]{#1}%
\providecommand \enquote  [1]{``#1''}%
\providecommand \bibnamefont  [1]{#1}%
\providecommand \bibfnamefont [1]{#1}%
\providecommand \citenamefont [1]{#1}%
\providecommand \href@noop [0]{\@secondoftwo}%
\providecommand \href [0]{\begingroup \@sanitize@url \@href}%
\providecommand \@href[1]{\@@startlink{#1}\@@href}%
\providecommand \@@href[1]{\endgroup#1\@@endlink}%
\providecommand \@sanitize@url [0]{\catcode `\\12\catcode `\$12\catcode
  `\&12\catcode `\#12\catcode `\^12\catcode `\_12\catcode `\%12\relax}%
\providecommand \@@startlink[1]{}%
\providecommand \@@endlink[0]{}%
\providecommand \url  [0]{\begingroup\@sanitize@url \@url }%
\providecommand \@url [1]{\endgroup\@href {#1}{\urlprefix }}%
\providecommand \urlprefix  [0]{URL }%
\providecommand \Eprint [0]{\href }%
\providecommand \doibase [0]{https://doi.org/}%
\providecommand \selectlanguage [0]{\@gobble}%
\providecommand \bibinfo  [0]{\@secondoftwo}%
\providecommand \bibfield  [0]{\@secondoftwo}%
\providecommand \translation [1]{[#1]}%
\providecommand \BibitemOpen [0]{}%
\providecommand \bibitemStop [0]{}%
\providecommand \bibitemNoStop [0]{.\EOS\space}%
\providecommand \EOS [0]{\spacefactor3000\relax}%
\providecommand \BibitemShut  [1]{\csname bibitem#1\endcsname}%
\let\auto@bib@innerbib\@empty
%</preamble>
\bibitem [{\citenamefont {Yeh}\ and\ \citenamefont
  {Lin}(2018)}]{yeh2018breakthrough}%
  \BibitemOpen
  \bibfield  {author} {\bibinfo {author} {\bibfnamefont {J.-W.}\ \bibnamefont
  {Yeh}}\ and\ \bibinfo {author} {\bibfnamefont {S.-J.}\ \bibnamefont {Lin}},\
  }\bibfield  {title} {\bibinfo {title} {Breakthrough applications of
  high-entropy materials},\ }\href@noop {} {\bibfield  {journal} {\bibinfo
  {journal} {Journal of Materials Research}\ }\textbf {\bibinfo {volume}
  {33}},\ \bibinfo {pages} {3129} (\bibinfo {year} {2018})}\BibitemShut
  {NoStop}%
\bibitem [{\citenamefont {Miracle}\ and\ \citenamefont
  {Senkov}(2017)}]{miracle2017critical}%
  \BibitemOpen
  \bibfield  {author} {\bibinfo {author} {\bibfnamefont {D.~B.}\ \bibnamefont
  {Miracle}}\ and\ \bibinfo {author} {\bibfnamefont {O.~N.}\ \bibnamefont
  {Senkov}},\ }\bibfield  {title} {\bibinfo {title} {A critical review of high
  entropy alloys and related concepts},\ }\href@noop {} {\bibfield  {journal}
  {\bibinfo  {journal} {Acta Materialia}\ }\textbf {\bibinfo {volume} {122}},\
  \bibinfo {pages} {448} (\bibinfo {year} {2017})}\BibitemShut {NoStop}%
\bibitem [{\citenamefont {George}\ \emph {et~al.}(2019)\citenamefont {George},
  \citenamefont {Raabe},\ and\ \citenamefont {Ritchie}}]{george2019high}%
  \BibitemOpen
  \bibfield  {author} {\bibinfo {author} {\bibfnamefont {E.~P.}\ \bibnamefont
  {George}}, \bibinfo {author} {\bibfnamefont {D.}~\bibnamefont {Raabe}},\ and\
  \bibinfo {author} {\bibfnamefont {R.~O.}\ \bibnamefont {Ritchie}},\
  }\bibfield  {title} {\bibinfo {title} {High-entropy alloys},\ }\href@noop {}
  {\bibfield  {journal} {\bibinfo  {journal} {Nature reviews materials}\
  }\textbf {\bibinfo {volume} {4}},\ \bibinfo {pages} {515} (\bibinfo {year}
  {2019})}\BibitemShut {NoStop}%
\bibitem [{\citenamefont {Ma}\ \emph {et~al.}(2021)\citenamefont {Ma},
  \citenamefont {Ma}, \citenamefont {Wang}, \citenamefont {Schweidler},
  \citenamefont {Botros}, \citenamefont {Fu}, \citenamefont {Hahn},
  \citenamefont {Brezesinski},\ and\ \citenamefont {Breitung}}]{ma2021high}%
  \BibitemOpen
  \bibfield  {author} {\bibinfo {author} {\bibfnamefont {Y.}~\bibnamefont
  {Ma}}, \bibinfo {author} {\bibfnamefont {Y.}~\bibnamefont {Ma}}, \bibinfo
  {author} {\bibfnamefont {Q.}~\bibnamefont {Wang}}, \bibinfo {author}
  {\bibfnamefont {S.}~\bibnamefont {Schweidler}}, \bibinfo {author}
  {\bibfnamefont {M.}~\bibnamefont {Botros}}, \bibinfo {author} {\bibfnamefont
  {T.}~\bibnamefont {Fu}}, \bibinfo {author} {\bibfnamefont {H.}~\bibnamefont
  {Hahn}}, \bibinfo {author} {\bibfnamefont {T.}~\bibnamefont {Brezesinski}},\
  and\ \bibinfo {author} {\bibfnamefont {B.}~\bibnamefont {Breitung}},\
  }\bibfield  {title} {\bibinfo {title} {High-entropy energy materials:
  challenges and new opportunities},\ }\href@noop {} {\bibfield  {journal}
  {\bibinfo  {journal} {Energy \& Environmental Science}\ }\textbf {\bibinfo
  {volume} {14}},\ \bibinfo {pages} {2883} (\bibinfo {year}
  {2021})}\BibitemShut {NoStop}%
\bibitem [{\citenamefont {Amiri}\ and\ \citenamefont
  {Shahbazian-Yassar}(2021)}]{amiri2021recent}%
  \BibitemOpen
  \bibfield  {author} {\bibinfo {author} {\bibfnamefont {A.}~\bibnamefont
  {Amiri}}\ and\ \bibinfo {author} {\bibfnamefont {R.}~\bibnamefont
  {Shahbazian-Yassar}},\ }\bibfield  {title} {\bibinfo {title} {Recent progress
  of high-entropy materials for energy storage and conversion},\ }\href@noop {}
  {\bibfield  {journal} {\bibinfo  {journal} {Journal of Materials Chemistry
  A}\ }\textbf {\bibinfo {volume} {9}},\ \bibinfo {pages} {782} (\bibinfo
  {year} {2021})}\BibitemShut {NoStop}%
\bibitem [{\citenamefont {Sun}\ and\ \citenamefont {Dai}(2021)}]{sun2021high}%
  \BibitemOpen
  \bibfield  {author} {\bibinfo {author} {\bibfnamefont {Y.}~\bibnamefont
  {Sun}}\ and\ \bibinfo {author} {\bibfnamefont {S.}~\bibnamefont {Dai}},\
  }\bibfield  {title} {\bibinfo {title} {High-entropy materials for catalysis:
  A new frontier},\ }\href@noop {} {\bibfield  {journal} {\bibinfo  {journal}
  {Science Advances}\ }\textbf {\bibinfo {volume} {7}},\ \bibinfo {pages}
  {eabg1600} (\bibinfo {year} {2021})}\BibitemShut {NoStop}%
\bibitem [{\citenamefont {Rickman}\ \emph {et~al.}(2020)\citenamefont
  {Rickman}, \citenamefont {Balasubramanian}, \citenamefont {Marvel},
  \citenamefont {Chan},\ and\ \citenamefont {Burton}}]{rickman2020machine}%
  \BibitemOpen
  \bibfield  {author} {\bibinfo {author} {\bibfnamefont {J.}~\bibnamefont
  {Rickman}}, \bibinfo {author} {\bibfnamefont {G.}~\bibnamefont
  {Balasubramanian}}, \bibinfo {author} {\bibfnamefont {C.}~\bibnamefont
  {Marvel}}, \bibinfo {author} {\bibfnamefont {H.}~\bibnamefont {Chan}},\ and\
  \bibinfo {author} {\bibfnamefont {M.-T.}\ \bibnamefont {Burton}},\ }\bibfield
   {title} {\bibinfo {title} {Machine learning strategies for high-entropy
  alloys},\ }\href@noop {} {\bibfield  {journal} {\bibinfo  {journal} {Journal
  of Applied Physics}\ }\textbf {\bibinfo {volume} {128}} (\bibinfo {year}
  {2020})}\BibitemShut {NoStop}%
\bibitem [{\citenamefont {Qiao}\ \emph {et~al.}(2021)\citenamefont {Qiao},
  \citenamefont {Liu},\ and\ \citenamefont {Zhu}}]{qiao2021focused}%
  \BibitemOpen
  \bibfield  {author} {\bibinfo {author} {\bibfnamefont {L.}~\bibnamefont
  {Qiao}}, \bibinfo {author} {\bibfnamefont {Y.}~\bibnamefont {Liu}},\ and\
  \bibinfo {author} {\bibfnamefont {J.}~\bibnamefont {Zhu}},\ }\bibfield
  {title} {\bibinfo {title} {A focused review on machine learning aided
  high-throughput methods in high entropy alloy},\ }\href@noop {} {\bibfield
  {journal} {\bibinfo  {journal} {Journal of Alloys and Compounds}\ }\textbf
  {\bibinfo {volume} {877}},\ \bibinfo {pages} {160295} (\bibinfo {year}
  {2021})}\BibitemShut {NoStop}%
\bibitem [{\citenamefont {Chen}\ \emph {et~al.}(2022)\citenamefont {Chen},
  \citenamefont {Chen}, \citenamefont {Gariepy}, \citenamefont {Yao},\ and\
  \citenamefont {Singh}}]{chen2022high}%
  \BibitemOpen
  \bibfield  {author} {\bibinfo {author} {\bibfnamefont {Z.~W.}\ \bibnamefont
  {Chen}}, \bibinfo {author} {\bibfnamefont {L.}~\bibnamefont {Chen}}, \bibinfo
  {author} {\bibfnamefont {Z.}~\bibnamefont {Gariepy}}, \bibinfo {author}
  {\bibfnamefont {X.}~\bibnamefont {Yao}},\ and\ \bibinfo {author}
  {\bibfnamefont {C.~V.}\ \bibnamefont {Singh}},\ }\bibfield  {title} {\bibinfo
  {title} {High-throughput and machine-learning accelerated design of high
  entropy alloy catalysts},\ }\href@noop {} {\bibfield  {journal} {\bibinfo
  {journal} {Trends in Chemistry}\ } (\bibinfo {year} {2022})}\BibitemShut
  {NoStop}%
\bibitem [{\citenamefont {Li}\ \emph {et~al.}(2020)\citenamefont {Li},
  \citenamefont {Xie}, \citenamefont {Wang}, \citenamefont {Liaw},\ and\
  \citenamefont {Zhang}}]{li2020high}%
  \BibitemOpen
  \bibfield  {author} {\bibinfo {author} {\bibfnamefont {R.}~\bibnamefont
  {Li}}, \bibinfo {author} {\bibfnamefont {L.}~\bibnamefont {Xie}}, \bibinfo
  {author} {\bibfnamefont {W.~Y.}\ \bibnamefont {Wang}}, \bibinfo {author}
  {\bibfnamefont {P.~K.}\ \bibnamefont {Liaw}},\ and\ \bibinfo {author}
  {\bibfnamefont {Y.}~\bibnamefont {Zhang}},\ }\bibfield  {title} {\bibinfo
  {title} {High-throughput calculations for high-entropy alloys: a brief
  review},\ }\href@noop {} {\bibfield  {journal} {\bibinfo  {journal}
  {Frontiers in Materials}\ }\textbf {\bibinfo {volume} {7}},\ \bibinfo {pages}
  {290} (\bibinfo {year} {2020})}\BibitemShut {NoStop}%
\bibitem [{\citenamefont {Huang}\ \emph {et~al.}(2019)\citenamefont {Huang},
  \citenamefont {Martin},\ and\ \citenamefont {Zhuang}}]{huang2019machine}%
  \BibitemOpen
  \bibfield  {author} {\bibinfo {author} {\bibfnamefont {W.}~\bibnamefont
  {Huang}}, \bibinfo {author} {\bibfnamefont {P.}~\bibnamefont {Martin}},\ and\
  \bibinfo {author} {\bibfnamefont {H.~L.}\ \bibnamefont {Zhuang}},\ }\bibfield
   {title} {\bibinfo {title} {Machine-learning phase prediction of high-entropy
  alloys},\ }\href@noop {} {\bibfield  {journal} {\bibinfo  {journal} {Acta
  Materialia}\ }\textbf {\bibinfo {volume} {169}},\ \bibinfo {pages} {225}
  (\bibinfo {year} {2019})}\BibitemShut {NoStop}%
\bibitem [{\citenamefont {Kaufmann}\ and\ \citenamefont
  {Vecchio}(2020)}]{kaufmann2020searching}%
  \BibitemOpen
  \bibfield  {author} {\bibinfo {author} {\bibfnamefont {K.}~\bibnamefont
  {Kaufmann}}\ and\ \bibinfo {author} {\bibfnamefont {K.~S.}\ \bibnamefont
  {Vecchio}},\ }\bibfield  {title} {\bibinfo {title} {Searching for high
  entropy alloys: A machine learning approach},\ }\href@noop {} {\bibfield
  {journal} {\bibinfo  {journal} {Acta Materialia}\ }\textbf {\bibinfo {volume}
  {198}},\ \bibinfo {pages} {178} (\bibinfo {year} {2020})}\BibitemShut
  {NoStop}%
\bibitem [{\citenamefont {Zhou}\ \emph {et~al.}(2023)\citenamefont {Zhou},
  \citenamefont {Xu}, \citenamefont {Gao}, \citenamefont {Zhang}, \citenamefont
  {Shi}, \citenamefont {Yuen},\ and\ \citenamefont {Zuo}}]{zhou2023machine}%
  \BibitemOpen
  \bibfield  {author} {\bibinfo {author} {\bibfnamefont {Q.}~\bibnamefont
  {Zhou}}, \bibinfo {author} {\bibfnamefont {F.}~\bibnamefont {Xu}}, \bibinfo
  {author} {\bibfnamefont {C.}~\bibnamefont {Gao}}, \bibinfo {author}
  {\bibfnamefont {D.}~\bibnamefont {Zhang}}, \bibinfo {author} {\bibfnamefont
  {X.}~\bibnamefont {Shi}}, \bibinfo {author} {\bibfnamefont {M.-F.}\
  \bibnamefont {Yuen}},\ and\ \bibinfo {author} {\bibfnamefont
  {D.}~\bibnamefont {Zuo}},\ }\bibfield  {title} {\bibinfo {title} {Machine
  learning-assisted mechanical property prediction and descriptor-property
  correlation analysis of high-entropy ceramics},\ }\href@noop {} {\bibfield
  {journal} {\bibinfo  {journal} {Ceramics International}\ }\textbf {\bibinfo
  {volume} {49}},\ \bibinfo {pages} {5760} (\bibinfo {year}
  {2023})}\BibitemShut {NoStop}%
\bibitem [{\citenamefont {Curtarolo}\ \emph {et~al.}(2012)\citenamefont
  {Curtarolo}, \citenamefont {Setyawan}, \citenamefont {Hart}, \citenamefont
  {Jahnatek}, \citenamefont {Chepulskii}, \citenamefont {Taylor}, \citenamefont
  {Wang}, \citenamefont {Xue}, \citenamefont {Yang}, \citenamefont {Levy} \emph
  {et~al.}}]{curtarolo2012aflow}%
  \BibitemOpen
  \bibfield  {author} {\bibinfo {author} {\bibfnamefont {S.}~\bibnamefont
  {Curtarolo}}, \bibinfo {author} {\bibfnamefont {W.}~\bibnamefont {Setyawan}},
  \bibinfo {author} {\bibfnamefont {G.~L.}\ \bibnamefont {Hart}}, \bibinfo
  {author} {\bibfnamefont {M.}~\bibnamefont {Jahnatek}}, \bibinfo {author}
  {\bibfnamefont {R.~V.}\ \bibnamefont {Chepulskii}}, \bibinfo {author}
  {\bibfnamefont {R.~H.}\ \bibnamefont {Taylor}}, \bibinfo {author}
  {\bibfnamefont {S.}~\bibnamefont {Wang}}, \bibinfo {author} {\bibfnamefont
  {J.}~\bibnamefont {Xue}}, \bibinfo {author} {\bibfnamefont {K.}~\bibnamefont
  {Yang}}, \bibinfo {author} {\bibfnamefont {O.}~\bibnamefont {Levy}}, \emph
  {et~al.},\ }\bibfield  {title} {\bibinfo {title} {Aflow: An automatic
  framework for high-throughput materials discovery},\ }\href@noop {}
  {\bibfield  {journal} {\bibinfo  {journal} {Computational Materials Science}\
  }\textbf {\bibinfo {volume} {58}},\ \bibinfo {pages} {218} (\bibinfo {year}
  {2012})}\BibitemShut {NoStop}%
\bibitem [{\citenamefont {Jain}\ \emph {et~al.}(2013)\citenamefont {Jain},
  \citenamefont {Ong}, \citenamefont {Hautier}, \citenamefont {Chen},
  \citenamefont {Richards}, \citenamefont {Dacek}, \citenamefont {Cholia},
  \citenamefont {Gunter}, \citenamefont {Skinner}, \citenamefont {Ceder} \emph
  {et~al.}}]{jain2013commentary}%
  \BibitemOpen
  \bibfield  {author} {\bibinfo {author} {\bibfnamefont {A.}~\bibnamefont
  {Jain}}, \bibinfo {author} {\bibfnamefont {S.~P.}\ \bibnamefont {Ong}},
  \bibinfo {author} {\bibfnamefont {G.}~\bibnamefont {Hautier}}, \bibinfo
  {author} {\bibfnamefont {W.}~\bibnamefont {Chen}}, \bibinfo {author}
  {\bibfnamefont {W.~D.}\ \bibnamefont {Richards}}, \bibinfo {author}
  {\bibfnamefont {S.}~\bibnamefont {Dacek}}, \bibinfo {author} {\bibfnamefont
  {S.}~\bibnamefont {Cholia}}, \bibinfo {author} {\bibfnamefont
  {D.}~\bibnamefont {Gunter}}, \bibinfo {author} {\bibfnamefont
  {D.}~\bibnamefont {Skinner}}, \bibinfo {author} {\bibfnamefont
  {G.}~\bibnamefont {Ceder}}, \emph {et~al.},\ }\bibfield  {title} {\bibinfo
  {title} {Commentary: The materials project: A materials genome approach to
  accelerating materials innovation},\ }\href@noop {} {\bibfield  {journal}
  {\bibinfo  {journal} {APL materials}\ }\textbf {\bibinfo {volume} {1}}
  (\bibinfo {year} {2013})}\BibitemShut {NoStop}%
\bibitem [{\citenamefont {Saal}\ \emph {et~al.}(2013)\citenamefont {Saal},
  \citenamefont {Kirklin}, \citenamefont {Aykol}, \citenamefont {Meredig},\
  and\ \citenamefont {Wolverton}}]{Saal2013}%
  \BibitemOpen
  \bibfield  {author} {\bibinfo {author} {\bibfnamefont {J.~E.}\ \bibnamefont
  {Saal}}, \bibinfo {author} {\bibfnamefont {S.}~\bibnamefont {Kirklin}},
  \bibinfo {author} {\bibfnamefont {M.}~\bibnamefont {Aykol}}, \bibinfo
  {author} {\bibfnamefont {B.}~\bibnamefont {Meredig}},\ and\ \bibinfo {author}
  {\bibfnamefont {C.}~\bibnamefont {Wolverton}},\ }\bibfield  {title} {\bibinfo
  {title} {{Materials design and discovery with high-throughput density
  functional theory: The open quantum materials database (OQMD)}},\ }\href
  {https://doi.org/10.1007/s11837-013-0755-4} {\bibfield  {journal} {\bibinfo
  {journal} {Jom}\ }\textbf {\bibinfo {volume} {65}},\ \bibinfo {pages} {1501}
  (\bibinfo {year} {2013})}\BibitemShut {NoStop}%
\bibitem [{\citenamefont {Choudhary}\ \emph {et~al.}(2020)\citenamefont
  {Choudhary}, \citenamefont {Garrity}, \citenamefont {Reid}, \citenamefont
  {DeCost}, \citenamefont {Biacchi}, \citenamefont {Hight~Walker},
  \citenamefont {Trautt}, \citenamefont {Hattrick-Simpers}, \citenamefont
  {Kusne}, \citenamefont {Centrone} \emph {et~al.}}]{choudhary2020joint}%
  \BibitemOpen
  \bibfield  {author} {\bibinfo {author} {\bibfnamefont {K.}~\bibnamefont
  {Choudhary}}, \bibinfo {author} {\bibfnamefont {K.~F.}\ \bibnamefont
  {Garrity}}, \bibinfo {author} {\bibfnamefont {A.~C.}\ \bibnamefont {Reid}},
  \bibinfo {author} {\bibfnamefont {B.}~\bibnamefont {DeCost}}, \bibinfo
  {author} {\bibfnamefont {A.~J.}\ \bibnamefont {Biacchi}}, \bibinfo {author}
  {\bibfnamefont {A.~R.}\ \bibnamefont {Hight~Walker}}, \bibinfo {author}
  {\bibfnamefont {Z.}~\bibnamefont {Trautt}}, \bibinfo {author} {\bibfnamefont
  {J.}~\bibnamefont {Hattrick-Simpers}}, \bibinfo {author} {\bibfnamefont
  {A.~G.}\ \bibnamefont {Kusne}}, \bibinfo {author} {\bibfnamefont
  {A.}~\bibnamefont {Centrone}}, \emph {et~al.},\ }\bibfield  {title} {\bibinfo
  {title} {The joint automated repository for various integrated simulations
  (jarvis) for data-driven materials design},\ }\href@noop {} {\bibfield
  {journal} {\bibinfo  {journal} {npj computational materials}\ }\textbf
  {\bibinfo {volume} {6}},\ \bibinfo {pages} {173} (\bibinfo {year}
  {2020})}\BibitemShut {NoStop}%
\bibitem [{\citenamefont {Chanussot}\ \emph {et~al.}(2021)\citenamefont
  {Chanussot}, \citenamefont {Das}, \citenamefont {Goyal}, \citenamefont
  {Lavril}, \citenamefont {Shuaibi}, \citenamefont {Riviere}, \citenamefont
  {Tran}, \citenamefont {Heras-Domingo}, \citenamefont {Ho}, \citenamefont
  {Hu}, \citenamefont {Palizhati}, \citenamefont {Sriram}, \citenamefont
  {Wood}, \citenamefont {Yoon}, \citenamefont {Parikh}, \citenamefont
  {Zitnick},\ and\ \citenamefont {Ulissi}}]{Chanussot2021}%
  \BibitemOpen
  \bibfield  {author} {\bibinfo {author} {\bibfnamefont {L.}~\bibnamefont
  {Chanussot}}, \bibinfo {author} {\bibfnamefont {A.}~\bibnamefont {Das}},
  \bibinfo {author} {\bibfnamefont {S.}~\bibnamefont {Goyal}}, \bibinfo
  {author} {\bibfnamefont {T.}~\bibnamefont {Lavril}}, \bibinfo {author}
  {\bibfnamefont {M.}~\bibnamefont {Shuaibi}}, \bibinfo {author} {\bibfnamefont
  {M.}~\bibnamefont {Riviere}}, \bibinfo {author} {\bibfnamefont
  {K.}~\bibnamefont {Tran}}, \bibinfo {author} {\bibfnamefont {J.}~\bibnamefont
  {Heras-Domingo}}, \bibinfo {author} {\bibfnamefont {C.}~\bibnamefont {Ho}},
  \bibinfo {author} {\bibfnamefont {W.}~\bibnamefont {Hu}}, \bibinfo {author}
  {\bibfnamefont {A.}~\bibnamefont {Palizhati}}, \bibinfo {author}
  {\bibfnamefont {A.}~\bibnamefont {Sriram}}, \bibinfo {author} {\bibfnamefont
  {B.}~\bibnamefont {Wood}}, \bibinfo {author} {\bibfnamefont {J.}~\bibnamefont
  {Yoon}}, \bibinfo {author} {\bibfnamefont {D.}~\bibnamefont {Parikh}},
  \bibinfo {author} {\bibfnamefont {C.~L.}\ \bibnamefont {Zitnick}},\ and\
  \bibinfo {author} {\bibfnamefont {Z.}~\bibnamefont {Ulissi}},\ }\bibfield
  {title} {\bibinfo {title} {{Open Catalyst 2020 (OC20) Dataset and Community
  Challenges}},\ }\href {https://doi.org/10.1021/acscatal.0c04525} {\bibfield
  {journal} {\bibinfo  {journal} {ACS Catal.}\ }\textbf {\bibinfo {volume}
  {11}},\ \bibinfo {pages} {6059} (\bibinfo {year} {2021})},\ \Eprint
  {https://arxiv.org/abs/2010.09990} {2010.09990} \BibitemShut {NoStop}%
\bibitem [{\citenamefont {Ikeda}\ \emph {et~al.}(2019)\citenamefont {Ikeda},
  \citenamefont {Grabowski},\ and\ \citenamefont {K{\"o}rmann}}]{ikeda2019ab}%
  \BibitemOpen
  \bibfield  {author} {\bibinfo {author} {\bibfnamefont {Y.}~\bibnamefont
  {Ikeda}}, \bibinfo {author} {\bibfnamefont {B.}~\bibnamefont {Grabowski}},\
  and\ \bibinfo {author} {\bibfnamefont {F.}~\bibnamefont {K{\"o}rmann}},\
  }\bibfield  {title} {\bibinfo {title} {Ab initio phase stabilities and
  mechanical properties of multicomponent alloys: A comprehensive review for
  high entropy alloys and compositionally complex alloys},\ }\href@noop {}
  {\bibfield  {journal} {\bibinfo  {journal} {Materials Characterization}\
  }\textbf {\bibinfo {volume} {147}},\ \bibinfo {pages} {464} (\bibinfo {year}
  {2019})}\BibitemShut {NoStop}%
\bibitem [{\citenamefont {Zunger}\ \emph {et~al.}(1990)\citenamefont {Zunger},
  \citenamefont {Wei}, \citenamefont {Ferreira},\ and\ \citenamefont
  {Bernard}}]{zunger1990special}%
  \BibitemOpen
  \bibfield  {author} {\bibinfo {author} {\bibfnamefont {A.}~\bibnamefont
  {Zunger}}, \bibinfo {author} {\bibfnamefont {S.-H.}\ \bibnamefont {Wei}},
  \bibinfo {author} {\bibfnamefont {L.}~\bibnamefont {Ferreira}},\ and\
  \bibinfo {author} {\bibfnamefont {J.~E.}\ \bibnamefont {Bernard}},\
  }\bibfield  {title} {\bibinfo {title} {Special quasirandom structures},\
  }\href@noop {} {\bibfield  {journal} {\bibinfo  {journal} {Physical review
  letters}\ }\textbf {\bibinfo {volume} {65}},\ \bibinfo {pages} {353}
  (\bibinfo {year} {1990})}\BibitemShut {NoStop}%
\bibitem [{\citenamefont {Van~de Walle}\ \emph {et~al.}(2013)\citenamefont
  {Van~de Walle}, \citenamefont {Tiwary}, \citenamefont {De~Jong},
  \citenamefont {Olmsted}, \citenamefont {Asta}, \citenamefont {Dick},
  \citenamefont {Shin}, \citenamefont {Wang}, \citenamefont {Chen},\ and\
  \citenamefont {Liu}}]{van2013efficient}%
  \BibitemOpen
  \bibfield  {author} {\bibinfo {author} {\bibfnamefont {A.}~\bibnamefont
  {Van~de Walle}}, \bibinfo {author} {\bibfnamefont {P.}~\bibnamefont
  {Tiwary}}, \bibinfo {author} {\bibfnamefont {M.}~\bibnamefont {De~Jong}},
  \bibinfo {author} {\bibfnamefont {D.}~\bibnamefont {Olmsted}}, \bibinfo
  {author} {\bibfnamefont {M.}~\bibnamefont {Asta}}, \bibinfo {author}
  {\bibfnamefont {A.}~\bibnamefont {Dick}}, \bibinfo {author} {\bibfnamefont
  {D.}~\bibnamefont {Shin}}, \bibinfo {author} {\bibfnamefont {Y.}~\bibnamefont
  {Wang}}, \bibinfo {author} {\bibfnamefont {L.-Q.}\ \bibnamefont {Chen}},\
  and\ \bibinfo {author} {\bibfnamefont {Z.-K.}\ \bibnamefont {Liu}},\
  }\bibfield  {title} {\bibinfo {title} {Efficient stochastic generation of
  special quasirandom structures},\ }\href@noop {} {\bibfield  {journal}
  {\bibinfo  {journal} {Calphad}\ }\textbf {\bibinfo {volume} {42}},\ \bibinfo
  {pages} {13} (\bibinfo {year} {2013})}\BibitemShut {NoStop}%
\bibitem [{\citenamefont {Yang}\ \emph {et~al.}(2023)\citenamefont {Yang},
  \citenamefont {Manganaris},\ and\ \citenamefont
  {Mannodi-Kanakkithodi}}]{yang2023high}%
  \BibitemOpen
  \bibfield  {author} {\bibinfo {author} {\bibfnamefont {J.}~\bibnamefont
  {Yang}}, \bibinfo {author} {\bibfnamefont {P.}~\bibnamefont {Manganaris}},\
  and\ \bibinfo {author} {\bibfnamefont {A.}~\bibnamefont
  {Mannodi-Kanakkithodi}},\ }\bibfield  {title} {\bibinfo {title} {A
  high-throughput computational dataset of halide perovskite alloys},\
  }\href@noop {} {\bibfield  {journal} {\bibinfo  {journal} {Digital
  Discovery}\ } (\bibinfo {year} {2023})}\BibitemShut {NoStop}%
\bibitem [{\citenamefont {Li}\ and\ \citenamefont {Fu}(2020)}]{li2020ground}%
  \BibitemOpen
  \bibfield  {author} {\bibinfo {author} {\bibfnamefont {K.}~\bibnamefont
  {Li}}\ and\ \bibinfo {author} {\bibfnamefont {C.-C.}\ \bibnamefont {Fu}},\
  }\bibfield  {title} {\bibinfo {title} {Ground-state properties and
  lattice-vibration effects of disordered fe-ni systems for phase stability
  predictions},\ }\href@noop {} {\bibfield  {journal} {\bibinfo  {journal}
  {Physical Review Materials}\ }\textbf {\bibinfo {volume} {4}},\ \bibinfo
  {pages} {023606} (\bibinfo {year} {2020})}\BibitemShut {NoStop}%
\bibitem [{\citenamefont {Shen}\ \emph {et~al.}(2022)\citenamefont {Shen},
  \citenamefont {Griesemer}, \citenamefont {Gopakumar}, \citenamefont
  {Baldassarri}, \citenamefont {Saal}, \citenamefont {Aykol}, \citenamefont
  {Hegde},\ and\ \citenamefont {Wolverton}}]{shen2022reflections}%
  \BibitemOpen
  \bibfield  {author} {\bibinfo {author} {\bibfnamefont {J.}~\bibnamefont
  {Shen}}, \bibinfo {author} {\bibfnamefont {S.~D.}\ \bibnamefont {Griesemer}},
  \bibinfo {author} {\bibfnamefont {A.}~\bibnamefont {Gopakumar}}, \bibinfo
  {author} {\bibfnamefont {B.}~\bibnamefont {Baldassarri}}, \bibinfo {author}
  {\bibfnamefont {J.~E.}\ \bibnamefont {Saal}}, \bibinfo {author}
  {\bibfnamefont {M.}~\bibnamefont {Aykol}}, \bibinfo {author} {\bibfnamefont
  {V.~I.}\ \bibnamefont {Hegde}},\ and\ \bibinfo {author} {\bibfnamefont
  {C.}~\bibnamefont {Wolverton}},\ }\bibfield  {title} {\bibinfo {title}
  {Reflections on one million compounds in the open quantum materials database
  (oqmd)},\ }\href@noop {} {\bibfield  {journal} {\bibinfo  {journal} {Journal
  of Physics: Materials}\ }\textbf {\bibinfo {volume} {5}},\ \bibinfo {pages}
  {031001} (\bibinfo {year} {2022})}\BibitemShut {NoStop}%
\bibitem [{\citenamefont {Bokas}\ \emph {et~al.}(2021)\citenamefont {Bokas},
  \citenamefont {Chen}, \citenamefont {Hilhorst}, \citenamefont {Jacques},
  \citenamefont {Gorsse},\ and\ \citenamefont {Hautier}}]{bokas2021unveiling}%
  \BibitemOpen
  \bibfield  {author} {\bibinfo {author} {\bibfnamefont {G.~B.}\ \bibnamefont
  {Bokas}}, \bibinfo {author} {\bibfnamefont {W.}~\bibnamefont {Chen}},
  \bibinfo {author} {\bibfnamefont {A.}~\bibnamefont {Hilhorst}}, \bibinfo
  {author} {\bibfnamefont {P.~J.}\ \bibnamefont {Jacques}}, \bibinfo {author}
  {\bibfnamefont {S.}~\bibnamefont {Gorsse}},\ and\ \bibinfo {author}
  {\bibfnamefont {G.}~\bibnamefont {Hautier}},\ }\bibfield  {title} {\bibinfo
  {title} {Unveiling the thermodynamic driving forces for high entropy alloys
  formation through big data ab initio analysis},\ }\href@noop {} {\bibfield
  {journal} {\bibinfo  {journal} {Scripta Materialia}\ }\textbf {\bibinfo
  {volume} {202}},\ \bibinfo {pages} {114000} (\bibinfo {year}
  {2021})}\BibitemShut {NoStop}%
\bibitem [{\citenamefont {Chen}\ \emph {et~al.}(2023)\citenamefont {Chen},
  \citenamefont {Hilhorst}, \citenamefont {Bokas}, \citenamefont {Gorsse},
  \citenamefont {Jacques},\ and\ \citenamefont {Hautier}}]{chen2023map}%
  \BibitemOpen
  \bibfield  {author} {\bibinfo {author} {\bibfnamefont {W.}~\bibnamefont
  {Chen}}, \bibinfo {author} {\bibfnamefont {A.}~\bibnamefont {Hilhorst}},
  \bibinfo {author} {\bibfnamefont {G.}~\bibnamefont {Bokas}}, \bibinfo
  {author} {\bibfnamefont {S.}~\bibnamefont {Gorsse}}, \bibinfo {author}
  {\bibfnamefont {P.~J.}\ \bibnamefont {Jacques}},\ and\ \bibinfo {author}
  {\bibfnamefont {G.}~\bibnamefont {Hautier}},\ }\bibfield  {title} {\bibinfo
  {title} {A map of single-phase high-entropy alloys},\ }\href@noop {}
  {\bibfield  {journal} {\bibinfo  {journal} {Nature Communications}\ }\textbf
  {\bibinfo {volume} {14}},\ \bibinfo {pages} {2856} (\bibinfo {year}
  {2023})}\BibitemShut {NoStop}%
\bibitem [{\citenamefont {Sarker}\ \emph {et~al.}(2018)\citenamefont {Sarker},
  \citenamefont {Harrington}, \citenamefont {Toher}, \citenamefont {Oses},
  \citenamefont {Samiee}, \citenamefont {Maria}, \citenamefont {Brenner},
  \citenamefont {Vecchio},\ and\ \citenamefont {Curtarolo}}]{sarker2018high}%
  \BibitemOpen
  \bibfield  {author} {\bibinfo {author} {\bibfnamefont {P.}~\bibnamefont
  {Sarker}}, \bibinfo {author} {\bibfnamefont {T.}~\bibnamefont {Harrington}},
  \bibinfo {author} {\bibfnamefont {C.}~\bibnamefont {Toher}}, \bibinfo
  {author} {\bibfnamefont {C.}~\bibnamefont {Oses}}, \bibinfo {author}
  {\bibfnamefont {M.}~\bibnamefont {Samiee}}, \bibinfo {author} {\bibfnamefont
  {J.-P.}\ \bibnamefont {Maria}}, \bibinfo {author} {\bibfnamefont {D.~W.}\
  \bibnamefont {Brenner}}, \bibinfo {author} {\bibfnamefont {K.~S.}\
  \bibnamefont {Vecchio}},\ and\ \bibinfo {author} {\bibfnamefont
  {S.}~\bibnamefont {Curtarolo}},\ }\bibfield  {title} {\bibinfo {title}
  {High-entropy high-hardness metal carbides discovered by entropy
  descriptors},\ }\href@noop {} {\bibfield  {journal} {\bibinfo  {journal}
  {Nature communications}\ }\textbf {\bibinfo {volume} {9}},\ \bibinfo {pages}
  {4980} (\bibinfo {year} {2018})}\BibitemShut {NoStop}%
\bibitem [{\citenamefont {Kaufmann}\ \emph {et~al.}(2020)\citenamefont
  {Kaufmann}, \citenamefont {Maryanovsky}, \citenamefont {Mellor},
  \citenamefont {Zhu}, \citenamefont {Rosengarten}, \citenamefont {Harrington},
  \citenamefont {Oses}, \citenamefont {Toher}, \citenamefont {Curtarolo},\ and\
  \citenamefont {Vecchio}}]{kaufmann2020discovery}%
  \BibitemOpen
  \bibfield  {author} {\bibinfo {author} {\bibfnamefont {K.}~\bibnamefont
  {Kaufmann}}, \bibinfo {author} {\bibfnamefont {D.}~\bibnamefont
  {Maryanovsky}}, \bibinfo {author} {\bibfnamefont {W.~M.}\ \bibnamefont
  {Mellor}}, \bibinfo {author} {\bibfnamefont {C.}~\bibnamefont {Zhu}},
  \bibinfo {author} {\bibfnamefont {A.~S.}\ \bibnamefont {Rosengarten}},
  \bibinfo {author} {\bibfnamefont {T.~J.}\ \bibnamefont {Harrington}},
  \bibinfo {author} {\bibfnamefont {C.}~\bibnamefont {Oses}}, \bibinfo {author}
  {\bibfnamefont {C.}~\bibnamefont {Toher}}, \bibinfo {author} {\bibfnamefont
  {S.}~\bibnamefont {Curtarolo}},\ and\ \bibinfo {author} {\bibfnamefont
  {K.~S.}\ \bibnamefont {Vecchio}},\ }\bibfield  {title} {\bibinfo {title}
  {Discovery of high-entropy ceramics via machine learning},\ }\href@noop {}
  {\bibfield  {journal} {\bibinfo  {journal} {Npj Computational Materials}\
  }\textbf {\bibinfo {volume} {6}},\ \bibinfo {pages} {42} (\bibinfo {year}
  {2020})}\BibitemShut {NoStop}%
\bibitem [{\citenamefont {Lederer}\ \emph {et~al.}(2018)\citenamefont
  {Lederer}, \citenamefont {Toher}, \citenamefont {Vecchio},\ and\
  \citenamefont {Curtarolo}}]{lederer2018search}%
  \BibitemOpen
  \bibfield  {author} {\bibinfo {author} {\bibfnamefont {Y.}~\bibnamefont
  {Lederer}}, \bibinfo {author} {\bibfnamefont {C.}~\bibnamefont {Toher}},
  \bibinfo {author} {\bibfnamefont {K.~S.}\ \bibnamefont {Vecchio}},\ and\
  \bibinfo {author} {\bibfnamefont {S.}~\bibnamefont {Curtarolo}},\ }\bibfield
  {title} {\bibinfo {title} {The search for high entropy alloys: a
  high-throughput ab-initio approach},\ }\href@noop {} {\bibfield  {journal}
  {\bibinfo  {journal} {Acta Materialia}\ }\textbf {\bibinfo {volume} {159}},\
  \bibinfo {pages} {364} (\bibinfo {year} {2018})}\BibitemShut {NoStop}%
\bibitem [{\citenamefont {Jiang}\ and\ \citenamefont
  {Uberuaga}(2016)}]{jiang2016efficient}%
  \BibitemOpen
  \bibfield  {author} {\bibinfo {author} {\bibfnamefont {C.}~\bibnamefont
  {Jiang}}\ and\ \bibinfo {author} {\bibfnamefont {B.~P.}\ \bibnamefont
  {Uberuaga}},\ }\bibfield  {title} {\bibinfo {title} {Efficient ab initio
  modeling of random multicomponent alloys},\ }\href@noop {} {\bibfield
  {journal} {\bibinfo  {journal} {Physical review letters}\ }\textbf {\bibinfo
  {volume} {116}},\ \bibinfo {pages} {105501} (\bibinfo {year}
  {2016})}\BibitemShut {NoStop}%
\bibitem [{\citenamefont {Sorkin}\ \emph {et~al.}(2022)\citenamefont {Sorkin},
  \citenamefont {Yu}, \citenamefont {Chen}, \citenamefont {Tan}, \citenamefont
  {Aitken},\ and\ \citenamefont {Zhang}}]{sorkin2022first}%
  \BibitemOpen
  \bibfield  {author} {\bibinfo {author} {\bibfnamefont {V.}~\bibnamefont
  {Sorkin}}, \bibinfo {author} {\bibfnamefont {Z.}~\bibnamefont {Yu}}, \bibinfo
  {author} {\bibfnamefont {S.}~\bibnamefont {Chen}}, \bibinfo {author}
  {\bibfnamefont {T.~L.}\ \bibnamefont {Tan}}, \bibinfo {author} {\bibfnamefont
  {Z.}~\bibnamefont {Aitken}},\ and\ \bibinfo {author} {\bibfnamefont
  {Y.}~\bibnamefont {Zhang}},\ }\bibfield  {title} {\bibinfo {title} {A
  first-principles-based high fidelity, high throughput approach for the design
  of high entropy alloys},\ }\href@noop {} {\bibfield  {journal} {\bibinfo
  {journal} {Scientific Reports}\ }\textbf {\bibinfo {volume} {12}},\ \bibinfo
  {pages} {11894} (\bibinfo {year} {2022})}\BibitemShut {NoStop}%
\bibitem [{\citenamefont {Vazquez}\ \emph {et~al.}(2022)\citenamefont
  {Vazquez}, \citenamefont {Singh}, \citenamefont {Sauceda}, \citenamefont
  {Couperthwaite}, \citenamefont {Britt}, \citenamefont {Youssef},
  \citenamefont {Johnson},\ and\ \citenamefont
  {Arr{\'o}yave}}]{vazquez2022efficient}%
  \BibitemOpen
  \bibfield  {author} {\bibinfo {author} {\bibfnamefont {G.}~\bibnamefont
  {Vazquez}}, \bibinfo {author} {\bibfnamefont {P.}~\bibnamefont {Singh}},
  \bibinfo {author} {\bibfnamefont {D.}~\bibnamefont {Sauceda}}, \bibinfo
  {author} {\bibfnamefont {R.}~\bibnamefont {Couperthwaite}}, \bibinfo {author}
  {\bibfnamefont {N.}~\bibnamefont {Britt}}, \bibinfo {author} {\bibfnamefont
  {K.}~\bibnamefont {Youssef}}, \bibinfo {author} {\bibfnamefont {D.~D.}\
  \bibnamefont {Johnson}},\ and\ \bibinfo {author} {\bibfnamefont
  {R.}~\bibnamefont {Arr{\'o}yave}},\ }\bibfield  {title} {\bibinfo {title}
  {Efficient machine-learning model for fast assessment of elastic properties
  of high-entropy alloys},\ }\href@noop {} {\bibfield  {journal} {\bibinfo
  {journal} {Acta Materialia}\ }\textbf {\bibinfo {volume} {232}},\ \bibinfo
  {pages} {117924} (\bibinfo {year} {2022})}\BibitemShut {NoStop}%
\bibitem [{\citenamefont {Zhang}\ \emph {et~al.}(2022)\citenamefont {Zhang},
  \citenamefont {Xu}, \citenamefont {Xiong}, \citenamefont {Ma}, \citenamefont
  {Wang}, \citenamefont {Wu},\ and\ \citenamefont {Zhao}}]{zhang2022design}%
  \BibitemOpen
  \bibfield  {author} {\bibinfo {author} {\bibfnamefont {J.}~\bibnamefont
  {Zhang}}, \bibinfo {author} {\bibfnamefont {B.}~\bibnamefont {Xu}}, \bibinfo
  {author} {\bibfnamefont {Y.}~\bibnamefont {Xiong}}, \bibinfo {author}
  {\bibfnamefont {S.}~\bibnamefont {Ma}}, \bibinfo {author} {\bibfnamefont
  {Z.}~\bibnamefont {Wang}}, \bibinfo {author} {\bibfnamefont {Z.}~\bibnamefont
  {Wu}},\ and\ \bibinfo {author} {\bibfnamefont {S.}~\bibnamefont {Zhao}},\
  }\bibfield  {title} {\bibinfo {title} {Design high-entropy carbide ceramics
  from machine learning},\ }\href@noop {} {\bibfield  {journal} {\bibinfo
  {journal} {npj Computational Materials}\ }\textbf {\bibinfo {volume} {8}},\
  \bibinfo {pages} {5} (\bibinfo {year} {2022})}\BibitemShut {NoStop}%
\bibitem [{\citenamefont {Li}\ \emph {et~al.}(2023{\natexlab{a}})\citenamefont
  {Li}, \citenamefont {DeCost}, \citenamefont {Choudhary}, \citenamefont
  {Greenwood},\ and\ \citenamefont {Hattrick-Simpers}}]{li2022critical}%
  \BibitemOpen
  \bibfield  {author} {\bibinfo {author} {\bibfnamefont {K.}~\bibnamefont
  {Li}}, \bibinfo {author} {\bibfnamefont {B.}~\bibnamefont {DeCost}}, \bibinfo
  {author} {\bibfnamefont {K.}~\bibnamefont {Choudhary}}, \bibinfo {author}
  {\bibfnamefont {M.}~\bibnamefont {Greenwood}},\ and\ \bibinfo {author}
  {\bibfnamefont {J.}~\bibnamefont {Hattrick-Simpers}},\ }\bibfield  {title}
  {\bibinfo {title} {A critical examination of robustness and generalizability
  of machine learning prediction of materials properties},\ }\href@noop {}
  {\bibfield  {journal} {\bibinfo  {journal} {npj Computational Materials}\
  }\textbf {\bibinfo {volume} {9}},\ \bibinfo {pages} {55} (\bibinfo {year}
  {2023}{\natexlab{a}})}\BibitemShut {NoStop}%
\bibitem [{\citenamefont {Li}\ \emph {et~al.}(2023{\natexlab{b}})\citenamefont
  {Li}, \citenamefont {Persaud}, \citenamefont {Choudhary}, \citenamefont
  {DeCost}, \citenamefont {Greenwood},\ and\ \citenamefont
  {Hattrick-Simpers}}]{li2023exploiting}%
  \BibitemOpen
  \bibfield  {author} {\bibinfo {author} {\bibfnamefont {K.}~\bibnamefont
  {Li}}, \bibinfo {author} {\bibfnamefont {D.}~\bibnamefont {Persaud}},
  \bibinfo {author} {\bibfnamefont {K.}~\bibnamefont {Choudhary}}, \bibinfo
  {author} {\bibfnamefont {B.}~\bibnamefont {DeCost}}, \bibinfo {author}
  {\bibfnamefont {M.}~\bibnamefont {Greenwood}},\ and\ \bibinfo {author}
  {\bibfnamefont {J.}~\bibnamefont {Hattrick-Simpers}},\ }\bibfield  {title}
  {\bibinfo {title} {Exploiting redundancy in large materials datasets for
  efficient machine learning with less data},\ }\href@noop {} {\bibfield
  {journal} {\bibinfo  {journal} {Nature Communications}\ }\textbf {\bibinfo
  {volume} {14}},\ \bibinfo {pages} {7283} (\bibinfo {year}
  {2023}{\natexlab{b}})}\BibitemShut {NoStop}%
\bibitem [{\citenamefont {Sch{\"u}tt}\ \emph {et~al.}(2014)\citenamefont
  {Sch{\"u}tt}, \citenamefont {Glawe}, \citenamefont {Brockherde},
  \citenamefont {Sanna}, \citenamefont {M{\"u}ller},\ and\ \citenamefont
  {Gross}}]{schutt2014represent}%
  \BibitemOpen
  \bibfield  {author} {\bibinfo {author} {\bibfnamefont {K.~T.}\ \bibnamefont
  {Sch{\"u}tt}}, \bibinfo {author} {\bibfnamefont {H.}~\bibnamefont {Glawe}},
  \bibinfo {author} {\bibfnamefont {F.}~\bibnamefont {Brockherde}}, \bibinfo
  {author} {\bibfnamefont {A.}~\bibnamefont {Sanna}}, \bibinfo {author}
  {\bibfnamefont {K.-R.}\ \bibnamefont {M{\"u}ller}},\ and\ \bibinfo {author}
  {\bibfnamefont {E.~K.}\ \bibnamefont {Gross}},\ }\bibfield  {title} {\bibinfo
  {title} {How to represent crystal structures for machine learning: Towards
  fast prediction of electronic properties},\ }\href@noop {} {\bibfield
  {journal} {\bibinfo  {journal} {Physical Review B}\ }\textbf {\bibinfo
  {volume} {89}},\ \bibinfo {pages} {205118} (\bibinfo {year}
  {2014})}\BibitemShut {NoStop}%
\bibitem [{\citenamefont {Faber}\ \emph {et~al.}(2015)\citenamefont {Faber},
  \citenamefont {Lindmaa}, \citenamefont {Von~Lilienfeld},\ and\ \citenamefont
  {Armiento}}]{faber2015crystal}%
  \BibitemOpen
  \bibfield  {author} {\bibinfo {author} {\bibfnamefont {F.}~\bibnamefont
  {Faber}}, \bibinfo {author} {\bibfnamefont {A.}~\bibnamefont {Lindmaa}},
  \bibinfo {author} {\bibfnamefont {O.~A.}\ \bibnamefont {Von~Lilienfeld}},\
  and\ \bibinfo {author} {\bibfnamefont {R.}~\bibnamefont {Armiento}},\
  }\bibfield  {title} {\bibinfo {title} {Crystal structure representations for
  machine learning models of formation energies},\ }\href@noop {} {\bibfield
  {journal} {\bibinfo  {journal} {International Journal of Quantum Chemistry}\
  }\textbf {\bibinfo {volume} {115}},\ \bibinfo {pages} {1094} (\bibinfo {year}
  {2015})}\BibitemShut {NoStop}%
\bibitem [{\citenamefont {Isayev}\ \emph {et~al.}(2017)\citenamefont {Isayev},
  \citenamefont {Oses}, \citenamefont {Toher}, \citenamefont {Gossett},
  \citenamefont {Curtarolo},\ and\ \citenamefont
  {Tropsha}}]{isayev2017universal}%
  \BibitemOpen
  \bibfield  {author} {\bibinfo {author} {\bibfnamefont {O.}~\bibnamefont
  {Isayev}}, \bibinfo {author} {\bibfnamefont {C.}~\bibnamefont {Oses}},
  \bibinfo {author} {\bibfnamefont {C.}~\bibnamefont {Toher}}, \bibinfo
  {author} {\bibfnamefont {E.}~\bibnamefont {Gossett}}, \bibinfo {author}
  {\bibfnamefont {S.}~\bibnamefont {Curtarolo}},\ and\ \bibinfo {author}
  {\bibfnamefont {A.}~\bibnamefont {Tropsha}},\ }\bibfield  {title} {\bibinfo
  {title} {Universal fragment descriptors for predicting properties of
  inorganic crystals},\ }\href@noop {} {\bibfield  {journal} {\bibinfo
  {journal} {Nature communications}\ }\textbf {\bibinfo {volume} {8}},\
  \bibinfo {pages} {15679} (\bibinfo {year} {2017})}\BibitemShut {NoStop}%
\bibitem [{\citenamefont {Ward}\ \emph {et~al.}(2017)\citenamefont {Ward},
  \citenamefont {Liu}, \citenamefont {Krishna}, \citenamefont {Hegde},
  \citenamefont {Agrawal}, \citenamefont {Choudhary},\ and\ \citenamefont
  {Wolverton}}]{ward2017including}%
  \BibitemOpen
  \bibfield  {author} {\bibinfo {author} {\bibfnamefont {L.}~\bibnamefont
  {Ward}}, \bibinfo {author} {\bibfnamefont {R.}~\bibnamefont {Liu}}, \bibinfo
  {author} {\bibfnamefont {A.}~\bibnamefont {Krishna}}, \bibinfo {author}
  {\bibfnamefont {V.~I.}\ \bibnamefont {Hegde}}, \bibinfo {author}
  {\bibfnamefont {A.}~\bibnamefont {Agrawal}}, \bibinfo {author} {\bibfnamefont
  {A.}~\bibnamefont {Choudhary}},\ and\ \bibinfo {author} {\bibfnamefont
  {C.}~\bibnamefont {Wolverton}},\ }\bibfield  {title} {\bibinfo {title}
  {Including crystal structure attributes in machine learning models of
  formation energies via voronoi tessellations},\ }\href@noop {} {\bibfield
  {journal} {\bibinfo  {journal} {Physical Review B}\ }\textbf {\bibinfo
  {volume} {96}},\ \bibinfo {pages} {024104} (\bibinfo {year}
  {2017})}\BibitemShut {NoStop}%
\bibitem [{\citenamefont {Ward}\ \emph {et~al.}(2018)\citenamefont {Ward},
  \citenamefont {Dunn}, \citenamefont {Faghaninia}, \citenamefont {Zimmermann},
  \citenamefont {Bajaj}, \citenamefont {Wang}, \citenamefont {Montoya},
  \citenamefont {Chen}, \citenamefont {Bystrom}, \citenamefont {Dylla} \emph
  {et~al.}}]{ward2018matminer}%
  \BibitemOpen
  \bibfield  {author} {\bibinfo {author} {\bibfnamefont {L.}~\bibnamefont
  {Ward}}, \bibinfo {author} {\bibfnamefont {A.}~\bibnamefont {Dunn}}, \bibinfo
  {author} {\bibfnamefont {A.}~\bibnamefont {Faghaninia}}, \bibinfo {author}
  {\bibfnamefont {N.~E.}\ \bibnamefont {Zimmermann}}, \bibinfo {author}
  {\bibfnamefont {S.}~\bibnamefont {Bajaj}}, \bibinfo {author} {\bibfnamefont
  {Q.}~\bibnamefont {Wang}}, \bibinfo {author} {\bibfnamefont {J.}~\bibnamefont
  {Montoya}}, \bibinfo {author} {\bibfnamefont {J.}~\bibnamefont {Chen}},
  \bibinfo {author} {\bibfnamefont {K.}~\bibnamefont {Bystrom}}, \bibinfo
  {author} {\bibfnamefont {M.}~\bibnamefont {Dylla}}, \emph {et~al.},\
  }\bibfield  {title} {\bibinfo {title} {Matminer: An open source toolkit for
  materials data mining},\ }\href@noop {} {\bibfield  {journal} {\bibinfo
  {journal} {Computational Materials Science}\ }\textbf {\bibinfo {volume}
  {152}},\ \bibinfo {pages} {60} (\bibinfo {year} {2018})}\BibitemShut
  {NoStop}%
\bibitem [{\citenamefont {Ong}\ \emph {et~al.}(2013)\citenamefont {Ong},
  \citenamefont {Richards}, \citenamefont {Jain}, \citenamefont {Hautier},
  \citenamefont {Kocher}, \citenamefont {Cholia}, \citenamefont {Gunter},
  \citenamefont {Chevrier}, \citenamefont {Persson},\ and\ \citenamefont
  {Ceder}}]{ong2013python}%
  \BibitemOpen
  \bibfield  {author} {\bibinfo {author} {\bibfnamefont {S.~P.}\ \bibnamefont
  {Ong}}, \bibinfo {author} {\bibfnamefont {W.~D.}\ \bibnamefont {Richards}},
  \bibinfo {author} {\bibfnamefont {A.}~\bibnamefont {Jain}}, \bibinfo {author}
  {\bibfnamefont {G.}~\bibnamefont {Hautier}}, \bibinfo {author} {\bibfnamefont
  {M.}~\bibnamefont {Kocher}}, \bibinfo {author} {\bibfnamefont
  {S.}~\bibnamefont {Cholia}}, \bibinfo {author} {\bibfnamefont
  {D.}~\bibnamefont {Gunter}}, \bibinfo {author} {\bibfnamefont {V.~L.}\
  \bibnamefont {Chevrier}}, \bibinfo {author} {\bibfnamefont {K.~A.}\
  \bibnamefont {Persson}},\ and\ \bibinfo {author} {\bibfnamefont
  {G.}~\bibnamefont {Ceder}},\ }\bibfield  {title} {\bibinfo {title} {Python
  materials genomics (pymatgen): A robust, open-source python library for
  materials analysis},\ }\href@noop {} {\bibfield  {journal} {\bibinfo
  {journal} {Computational Materials Science}\ }\textbf {\bibinfo {volume}
  {68}},\ \bibinfo {pages} {314} (\bibinfo {year} {2013})}\BibitemShut
  {NoStop}%
\bibitem [{\citenamefont {Choudhary}\ \emph {et~al.}(2022)\citenamefont
  {Choudhary}, \citenamefont {DeCost}, \citenamefont {Chen}, \citenamefont
  {Jain}, \citenamefont {Tavazza}, \citenamefont {Cohn}, \citenamefont {Park},
  \citenamefont {Choudhary}, \citenamefont {Agrawal}, \citenamefont {Billinge}
  \emph {et~al.}}]{choudhary2022recent}%
  \BibitemOpen
  \bibfield  {author} {\bibinfo {author} {\bibfnamefont {K.}~\bibnamefont
  {Choudhary}}, \bibinfo {author} {\bibfnamefont {B.}~\bibnamefont {DeCost}},
  \bibinfo {author} {\bibfnamefont {C.}~\bibnamefont {Chen}}, \bibinfo {author}
  {\bibfnamefont {A.}~\bibnamefont {Jain}}, \bibinfo {author} {\bibfnamefont
  {F.}~\bibnamefont {Tavazza}}, \bibinfo {author} {\bibfnamefont
  {R.}~\bibnamefont {Cohn}}, \bibinfo {author} {\bibfnamefont {C.~W.}\
  \bibnamefont {Park}}, \bibinfo {author} {\bibfnamefont {A.}~\bibnamefont
  {Choudhary}}, \bibinfo {author} {\bibfnamefont {A.}~\bibnamefont {Agrawal}},
  \bibinfo {author} {\bibfnamefont {S.~J.}\ \bibnamefont {Billinge}}, \emph
  {et~al.},\ }\bibfield  {title} {\bibinfo {title} {Recent advances and
  applications of deep learning methods in materials science},\ }\href@noop {}
  {\bibfield  {journal} {\bibinfo  {journal} {npj Computational Materials}\
  }\textbf {\bibinfo {volume} {8}},\ \bibinfo {pages} {59} (\bibinfo {year}
  {2022})}\BibitemShut {NoStop}%
\bibitem [{\citenamefont {Kingsbury}\ \emph {et~al.}(2022)\citenamefont
  {Kingsbury}, \citenamefont {Rosen}, \citenamefont {Gupta}, \citenamefont
  {Munro}, \citenamefont {Ong}, \citenamefont {Jain}, \citenamefont
  {Dwaraknath}, \citenamefont {Horton},\ and\ \citenamefont
  {Persson}}]{kingsbury2022flexible}%
  \BibitemOpen
  \bibfield  {author} {\bibinfo {author} {\bibfnamefont {R.~S.}\ \bibnamefont
  {Kingsbury}}, \bibinfo {author} {\bibfnamefont {A.~S.}\ \bibnamefont
  {Rosen}}, \bibinfo {author} {\bibfnamefont {A.~S.}\ \bibnamefont {Gupta}},
  \bibinfo {author} {\bibfnamefont {J.~M.}\ \bibnamefont {Munro}}, \bibinfo
  {author} {\bibfnamefont {S.~P.}\ \bibnamefont {Ong}}, \bibinfo {author}
  {\bibfnamefont {A.}~\bibnamefont {Jain}}, \bibinfo {author} {\bibfnamefont
  {S.}~\bibnamefont {Dwaraknath}}, \bibinfo {author} {\bibfnamefont {M.~K.}\
  \bibnamefont {Horton}},\ and\ \bibinfo {author} {\bibfnamefont {K.~A.}\
  \bibnamefont {Persson}},\ }\bibfield  {title} {\bibinfo {title} {A flexible
  and scalable scheme for mixing computed formation energies from different
  levels of theory},\ }\href@noop {} {\bibfield  {journal} {\bibinfo  {journal}
  {npj Computational Materials}\ }\textbf {\bibinfo {volume} {8}},\ \bibinfo
  {pages} {195} (\bibinfo {year} {2022})}\BibitemShut {NoStop}%
\bibitem [{\citenamefont {Stevanovi{\'c}}\ \emph {et~al.}(2012)\citenamefont
  {Stevanovi{\'c}}, \citenamefont {Lany}, \citenamefont {Zhang},\ and\
  \citenamefont {Zunger}}]{stevanovic2012correcting}%
  \BibitemOpen
  \bibfield  {author} {\bibinfo {author} {\bibfnamefont {V.}~\bibnamefont
  {Stevanovi{\'c}}}, \bibinfo {author} {\bibfnamefont {S.}~\bibnamefont
  {Lany}}, \bibinfo {author} {\bibfnamefont {X.}~\bibnamefont {Zhang}},\ and\
  \bibinfo {author} {\bibfnamefont {A.}~\bibnamefont {Zunger}},\ }\bibfield
  {title} {\bibinfo {title} {Correcting density functional theory for accurate
  predictions of compound enthalpies of formation: Fitted elemental-phase
  reference energies},\ }\href@noop {} {\bibfield  {journal} {\bibinfo
  {journal} {Physical Review B}\ }\textbf {\bibinfo {volume} {85}},\ \bibinfo
  {pages} {115104} (\bibinfo {year} {2012})}\BibitemShut {NoStop}%
\bibitem [{\citenamefont {Bartel}\ \emph {et~al.}(2019)\citenamefont {Bartel},
  \citenamefont {Weimer}, \citenamefont {Lany}, \citenamefont {Musgrave},\ and\
  \citenamefont {Holder}}]{bartel2019role}%
  \BibitemOpen
  \bibfield  {author} {\bibinfo {author} {\bibfnamefont {C.~J.}\ \bibnamefont
  {Bartel}}, \bibinfo {author} {\bibfnamefont {A.~W.}\ \bibnamefont {Weimer}},
  \bibinfo {author} {\bibfnamefont {S.}~\bibnamefont {Lany}}, \bibinfo {author}
  {\bibfnamefont {C.~B.}\ \bibnamefont {Musgrave}},\ and\ \bibinfo {author}
  {\bibfnamefont {A.~M.}\ \bibnamefont {Holder}},\ }\bibfield  {title}
  {\bibinfo {title} {The role of decomposition reactions in assessing
  first-principles predictions of solid stability},\ }\href@noop {} {\bibfield
  {journal} {\bibinfo  {journal} {npj Computational Materials}\ }\textbf
  {\bibinfo {volume} {5}},\ \bibinfo {pages} {4} (\bibinfo {year}
  {2019})}\BibitemShut {NoStop}%
\bibitem [{\citenamefont {Kirklin}\ \emph {et~al.}(2015)\citenamefont
  {Kirklin}, \citenamefont {Saal}, \citenamefont {Meredig}, \citenamefont
  {Thompson}, \citenamefont {Doak}, \citenamefont {Aykol}, \citenamefont
  {R{\"u}hl},\ and\ \citenamefont {Wolverton}}]{kirklin2015open}%
  \BibitemOpen
  \bibfield  {author} {\bibinfo {author} {\bibfnamefont {S.}~\bibnamefont
  {Kirklin}}, \bibinfo {author} {\bibfnamefont {J.~E.}\ \bibnamefont {Saal}},
  \bibinfo {author} {\bibfnamefont {B.}~\bibnamefont {Meredig}}, \bibinfo
  {author} {\bibfnamefont {A.}~\bibnamefont {Thompson}}, \bibinfo {author}
  {\bibfnamefont {J.~W.}\ \bibnamefont {Doak}}, \bibinfo {author}
  {\bibfnamefont {M.}~\bibnamefont {Aykol}}, \bibinfo {author} {\bibfnamefont
  {S.}~\bibnamefont {R{\"u}hl}},\ and\ \bibinfo {author} {\bibfnamefont
  {C.}~\bibnamefont {Wolverton}},\ }\bibfield  {title} {\bibinfo {title} {The
  open quantum materials database (oqmd): assessing the accuracy of dft
  formation energies},\ }\href@noop {} {\bibfield  {journal} {\bibinfo
  {journal} {npj Computational Materials}\ }\textbf {\bibinfo {volume} {1}},\
  \bibinfo {pages} {1} (\bibinfo {year} {2015})}\BibitemShut {NoStop}%
\bibitem [{\citenamefont {Bartel}(2022)}]{bartel2022review}%
  \BibitemOpen
  \bibfield  {author} {\bibinfo {author} {\bibfnamefont {C.~J.}\ \bibnamefont
  {Bartel}},\ }\bibfield  {title} {\bibinfo {title} {Review of computational
  approaches to predict the thermodynamic stability of inorganic solids},\
  }\href@noop {} {\bibfield  {journal} {\bibinfo  {journal} {Journal of
  Materials Science}\ }\textbf {\bibinfo {volume} {57}},\ \bibinfo {pages}
  {10475} (\bibinfo {year} {2022})}\BibitemShut {NoStop}%
\bibitem [{\citenamefont {Griesemer}\ \emph {et~al.}(2023)\citenamefont
  {Griesemer}, \citenamefont {Xia},\ and\ \citenamefont
  {Wolverton}}]{griesemer2023accelerating}%
  \BibitemOpen
  \bibfield  {author} {\bibinfo {author} {\bibfnamefont {S.~D.}\ \bibnamefont
  {Griesemer}}, \bibinfo {author} {\bibfnamefont {Y.}~\bibnamefont {Xia}},\
  and\ \bibinfo {author} {\bibfnamefont {C.}~\bibnamefont {Wolverton}},\
  }\bibfield  {title} {\bibinfo {title} {Accelerating the prediction of stable
  materials with machine learning},\ }\href@noop {} {\bibfield  {journal}
  {\bibinfo  {journal} {Nature Computational Science}\ }\textbf {\bibinfo
  {volume} {3}},\ \bibinfo {pages} {934} (\bibinfo {year} {2023})}\BibitemShut
  {NoStop}%
\bibitem [{\citenamefont {Gong}\ \emph
  {et~al.}(2022{\natexlab{a}})\citenamefont {Gong}, \citenamefont {Wang},
  \citenamefont {Xie}, \citenamefont {Chae}, \citenamefont {Liu}, \citenamefont
  {Shao-Horn},\ and\ \citenamefont {Grossman}}]{gong2022calibrating}%
  \BibitemOpen
  \bibfield  {author} {\bibinfo {author} {\bibfnamefont {S.}~\bibnamefont
  {Gong}}, \bibinfo {author} {\bibfnamefont {S.}~\bibnamefont {Wang}}, \bibinfo
  {author} {\bibfnamefont {T.}~\bibnamefont {Xie}}, \bibinfo {author}
  {\bibfnamefont {W.~H.}\ \bibnamefont {Chae}}, \bibinfo {author}
  {\bibfnamefont {R.}~\bibnamefont {Liu}}, \bibinfo {author} {\bibfnamefont
  {Y.}~\bibnamefont {Shao-Horn}},\ and\ \bibinfo {author} {\bibfnamefont
  {J.~C.}\ \bibnamefont {Grossman}},\ }\bibfield  {title} {\bibinfo {title}
  {Calibrating dft formation enthalpy calculations by multifidelity machine
  learning},\ }\href@noop {} {\bibfield  {journal} {\bibinfo  {journal} {JACS
  Au}\ }\textbf {\bibinfo {volume} {2}},\ \bibinfo {pages} {1964} (\bibinfo
  {year} {2022}{\natexlab{a}})}\BibitemShut {NoStop}%
\bibitem [{\citenamefont {Mao}\ \emph {et~al.}(2021)\citenamefont {Mao},
  \citenamefont {Yang}, \citenamefont {Sheng}, \citenamefont {Wang},
  \citenamefont {Ouyang}, \citenamefont {Ye}, \citenamefont {Yang},\ and\
  \citenamefont {Zhang}}]{mao2021prediction}%
  \BibitemOpen
  \bibfield  {author} {\bibinfo {author} {\bibfnamefont {Y.}~\bibnamefont
  {Mao}}, \bibinfo {author} {\bibfnamefont {H.}~\bibnamefont {Yang}}, \bibinfo
  {author} {\bibfnamefont {Y.}~\bibnamefont {Sheng}}, \bibinfo {author}
  {\bibfnamefont {J.}~\bibnamefont {Wang}}, \bibinfo {author} {\bibfnamefont
  {R.}~\bibnamefont {Ouyang}}, \bibinfo {author} {\bibfnamefont
  {C.}~\bibnamefont {Ye}}, \bibinfo {author} {\bibfnamefont {J.}~\bibnamefont
  {Yang}},\ and\ \bibinfo {author} {\bibfnamefont {W.}~\bibnamefont {Zhang}},\
  }\bibfield  {title} {\bibinfo {title} {Prediction and classification of
  formation energies of binary compounds by machine learning: an approach
  without crystal structure information},\ }\href@noop {} {\bibfield  {journal}
  {\bibinfo  {journal} {ACS omega}\ }\textbf {\bibinfo {volume} {6}},\ \bibinfo
  {pages} {14533} (\bibinfo {year} {2021})}\BibitemShut {NoStop}%
\bibitem [{\citenamefont {Kresse}\ and\ \citenamefont
  {Hafner}(1993)}]{Kresse1993a}%
  \BibitemOpen
  \bibfield  {author} {\bibinfo {author} {\bibfnamefont {G.}~\bibnamefont
  {Kresse}}\ and\ \bibinfo {author} {\bibfnamefont {J.}~\bibnamefont
  {Hafner}},\ }\bibfield  {title} {\bibinfo {title} {{Ab initio molecular
  dynamics for liquid metals}},\ }\href
  {https://doi.org/10.1103/PhysRevB.47.558} {\bibfield  {journal} {\bibinfo
  {journal} {Phys. Rev. B}\ }\textbf {\bibinfo {volume} {47}},\ \bibinfo
  {pages} {558} (\bibinfo {year} {1993})}\BibitemShut {NoStop}%
\bibitem [{\citenamefont {Kresse}\ and\ \citenamefont
  {Furthm{\"{u}}ller}(1996{\natexlab{a}})}]{Kresse1996e}%
  \BibitemOpen
  \bibfield  {author} {\bibinfo {author} {\bibfnamefont {G.}~\bibnamefont
  {Kresse}}\ and\ \bibinfo {author} {\bibfnamefont {J.}~\bibnamefont
  {Furthm{\"{u}}ller}},\ }\bibfield  {title} {\bibinfo {title} {{Efficiency of
  ab-initio total energy calculations for metals and semiconductors using a
  plane-wave basis set}},\ }\href
  {https://doi.org/10.1016/0927-0256(96)00008-0} {\bibfield  {journal}
  {\bibinfo  {journal} {Comput. Mater. Sci.}\ }\textbf {\bibinfo {volume}
  {6}},\ \bibinfo {pages} {15} (\bibinfo {year}
  {1996}{\natexlab{a}})}\BibitemShut {NoStop}%
\bibitem [{\citenamefont {Kresse}\ and\ \citenamefont
  {Furthm{\"{u}}ller}(1996{\natexlab{b}})}]{Kresse1996c}%
  \BibitemOpen
  \bibfield  {author} {\bibinfo {author} {\bibfnamefont {G.}~\bibnamefont
  {Kresse}}\ and\ \bibinfo {author} {\bibfnamefont {J.}~\bibnamefont
  {Furthm{\"{u}}ller}},\ }\bibfield  {title} {\bibinfo {title} {{Efficient
  iterative schemes for ab initio total-energy calculations using a plane-wave
  basis set}},\ }\href {https://doi.org/10.1103/PhysRevB.54.11169} {\bibfield
  {journal} {\bibinfo  {journal} {Phys. Rev. B}\ }\textbf {\bibinfo {volume}
  {54}},\ \bibinfo {pages} {11169} (\bibinfo {year}
  {1996}{\natexlab{b}})}\BibitemShut {NoStop}%
\bibitem [{\citenamefont {Perdew}\ \emph {et~al.}(1996)\citenamefont {Perdew},
  \citenamefont {Burke},\ and\ \citenamefont {Ernzerhof}}]{Perdew1996a}%
  \BibitemOpen
  \bibfield  {author} {\bibinfo {author} {\bibfnamefont {J.~P.}\ \bibnamefont
  {Perdew}}, \bibinfo {author} {\bibfnamefont {K.}~\bibnamefont {Burke}},\ and\
  \bibinfo {author} {\bibfnamefont {M.}~\bibnamefont {Ernzerhof}},\ }\bibfield
  {title} {\bibinfo {title} {{Generalized Gradient Approximation Made
  Simple}},\ }\href {https://doi.org/10.1103/PhysRevLett.77.3865} {\bibfield
  {journal} {\bibinfo  {journal} {Phys. Rev. Lett.}\ }\textbf {\bibinfo
  {volume} {77}},\ \bibinfo {pages} {3865} (\bibinfo {year}
  {1996})}\BibitemShut {NoStop}%
\bibitem [{\citenamefont {Methfessel}\ and\ \citenamefont
  {Paxton}(1989)}]{Methfessel1989}%
  \BibitemOpen
  \bibfield  {author} {\bibinfo {author} {\bibfnamefont {M.}~\bibnamefont
  {Methfessel}}\ and\ \bibinfo {author} {\bibfnamefont {A.~T.}\ \bibnamefont
  {Paxton}},\ }\bibfield  {title} {\bibinfo {title} {{High-precision sampling
  for Brillouin-zone integration in metals}},\ }\href
  {https://doi.org/10.1103/PhysRevB.40.3616} {\bibfield  {journal} {\bibinfo
  {journal} {Phys. Rev. B}\ }\textbf {\bibinfo {volume} {40}},\ \bibinfo
  {pages} {3616} (\bibinfo {year} {1989})}\BibitemShut {NoStop}%
\bibitem [{\citenamefont {Monkhorst}\ and\ \citenamefont
  {Pack}(1976)}]{Monkhorst1976a}%
  \BibitemOpen
  \bibfield  {author} {\bibinfo {author} {\bibfnamefont {H.~J.}\ \bibnamefont
  {Monkhorst}}\ and\ \bibinfo {author} {\bibfnamefont {J.~D.}\ \bibnamefont
  {Pack}},\ }\bibfield  {title} {\bibinfo {title} {{Special points for
  Brillouin-zone integrations}},\ }\href
  {https://doi.org/10.1103/PhysRevB.13.5188} {\bibfield  {journal} {\bibinfo
  {journal} {Phys. Rev. B}\ }\textbf {\bibinfo {volume} {13}},\ \bibinfo
  {pages} {5188} (\bibinfo {year} {1976})}\BibitemShut {NoStop}%
\bibitem [{\citenamefont {Van De~Walle}\ \emph {et~al.}(2002)\citenamefont {Van
  De~Walle}, \citenamefont {Asta},\ and\ \citenamefont {Ceder}}]{van2002alloy}%
  \BibitemOpen
  \bibfield  {author} {\bibinfo {author} {\bibfnamefont {A.}~\bibnamefont {Van
  De~Walle}}, \bibinfo {author} {\bibfnamefont {M.}~\bibnamefont {Asta}},\ and\
  \bibinfo {author} {\bibfnamefont {G.}~\bibnamefont {Ceder}},\ }\bibfield
  {title} {\bibinfo {title} {The alloy theoretic automated toolkit: A user
  guide},\ }\href@noop {} {\bibfield  {journal} {\bibinfo  {journal} {Calphad}\
  }\textbf {\bibinfo {volume} {26}},\ \bibinfo {pages} {539} (\bibinfo {year}
  {2002})}\BibitemShut {NoStop}%
\bibitem [{\citenamefont {Chen}\ and\ \citenamefont
  {Guestrin}(2016)}]{xgboost}%
  \BibitemOpen
  \bibfield  {author} {\bibinfo {author} {\bibfnamefont {T.}~\bibnamefont
  {Chen}}\ and\ \bibinfo {author} {\bibfnamefont {C.}~\bibnamefont
  {Guestrin}},\ }\bibfield  {title} {\bibinfo {title} {{XGBoost}},\ }in\ \href
  {https://doi.org/10.1145/2939672.2939785} {\emph {\bibinfo {booktitle} {Proc.
  22nd ACM SIGKDD Int. Conf. Knowl. Discov. Data Min.}}}\ (\bibinfo
  {publisher} {ACM},\ \bibinfo {address} {New York, NY, USA},\ \bibinfo {year}
  {2016})\ pp.\ \bibinfo {pages} {785--794}\BibitemShut {NoStop}%
\bibitem [{\citenamefont {Breiman}(2001)}]{breiman2001random}%
  \BibitemOpen
  \bibfield  {author} {\bibinfo {author} {\bibfnamefont {L.}~\bibnamefont
  {Breiman}},\ }\bibfield  {title} {\bibinfo {title} {Random forests},\
  }\href@noop {} {\bibfield  {journal} {\bibinfo  {journal} {Machine learning}\
  }\textbf {\bibinfo {volume} {45}},\ \bibinfo {pages} {5} (\bibinfo {year}
  {2001})}\BibitemShut {NoStop}%
\bibitem [{\citenamefont {Pedregosa}\ \emph {et~al.}(2011)\citenamefont
  {Pedregosa}, \citenamefont {Varoquaux}, \citenamefont {Gramfort},
  \citenamefont {Michel}, \citenamefont {Thirion}, \citenamefont {Grisel},
  \citenamefont {Blondel}, \citenamefont {Prettenhofer}, \citenamefont {Weiss},
  \citenamefont {Dubourg}, \citenamefont {Vanderplas}, \citenamefont {Passos},
  \citenamefont {Cournapeau}, \citenamefont {Brucher}, \citenamefont {Perrot},\
  and\ \citenamefont {Duchesnay}}]{sklearn}%
  \BibitemOpen
  \bibfield  {author} {\bibinfo {author} {\bibfnamefont {F.}~\bibnamefont
  {Pedregosa}}, \bibinfo {author} {\bibfnamefont {G.}~\bibnamefont
  {Varoquaux}}, \bibinfo {author} {\bibfnamefont {A.}~\bibnamefont {Gramfort}},
  \bibinfo {author} {\bibfnamefont {V.}~\bibnamefont {Michel}}, \bibinfo
  {author} {\bibfnamefont {B.}~\bibnamefont {Thirion}}, \bibinfo {author}
  {\bibfnamefont {O.}~\bibnamefont {Grisel}}, \bibinfo {author} {\bibfnamefont
  {M.}~\bibnamefont {Blondel}}, \bibinfo {author} {\bibfnamefont
  {P.}~\bibnamefont {Prettenhofer}}, \bibinfo {author} {\bibfnamefont
  {R.}~\bibnamefont {Weiss}}, \bibinfo {author} {\bibfnamefont
  {V.}~\bibnamefont {Dubourg}}, \bibinfo {author} {\bibfnamefont
  {J.}~\bibnamefont {Vanderplas}}, \bibinfo {author} {\bibfnamefont
  {A.}~\bibnamefont {Passos}}, \bibinfo {author} {\bibfnamefont
  {D.}~\bibnamefont {Cournapeau}}, \bibinfo {author} {\bibfnamefont
  {M.}~\bibnamefont {Brucher}}, \bibinfo {author} {\bibfnamefont
  {M.}~\bibnamefont {Perrot}},\ and\ \bibinfo {author} {\bibfnamefont
  {{\'{E}}.}~\bibnamefont {Duchesnay}},\ }\bibfield  {title} {\bibinfo {title}
  {{Scikit-learn: Machine Learning in Python}},\ }\href@noop {} {\bibfield
  {journal} {\bibinfo  {journal} {J. Mach. Learn. Res.}\ }\textbf {\bibinfo
  {volume} {12}},\ \bibinfo {pages} {2825} (\bibinfo {year}
  {2011})}\BibitemShut {NoStop}%
\bibitem [{\citenamefont {Choudhary}\ and\ \citenamefont
  {DeCost}(2021)}]{choudhary2021atomistic}%
  \BibitemOpen
  \bibfield  {author} {\bibinfo {author} {\bibfnamefont {K.}~\bibnamefont
  {Choudhary}}\ and\ \bibinfo {author} {\bibfnamefont {B.}~\bibnamefont
  {DeCost}},\ }\bibfield  {title} {\bibinfo {title} {Atomistic line graph
  neural network for improved materials property predictions},\ }\href@noop {}
  {\bibfield  {journal} {\bibinfo  {journal} {npj Computational Materials}\
  }\textbf {\bibinfo {volume} {7}},\ \bibinfo {pages} {185} (\bibinfo {year}
  {2021})}\BibitemShut {NoStop}%
\bibitem [{\citenamefont {Ward}\ \emph {et~al.}(2016)\citenamefont {Ward},
  \citenamefont {Agrawal}, \citenamefont {Choudhary},\ and\ \citenamefont
  {Wolverton}}]{Ward2016}%
  \BibitemOpen
  \bibfield  {author} {\bibinfo {author} {\bibfnamefont {L.}~\bibnamefont
  {Ward}}, \bibinfo {author} {\bibfnamefont {A.}~\bibnamefont {Agrawal}},
  \bibinfo {author} {\bibfnamefont {A.}~\bibnamefont {Choudhary}},\ and\
  \bibinfo {author} {\bibfnamefont {C.}~\bibnamefont {Wolverton}},\ }\bibfield
  {title} {\bibinfo {title} {{A general-purpose machine learning framework for
  predicting properties of inorganic materials}},\ }\href
  {https://doi.org/10.1038/npjcompumats.2016.28} {\bibfield  {journal}
  {\bibinfo  {journal} {npj Comput. Mater.}\ }\textbf {\bibinfo {volume} {2}},\
  \bibinfo {pages} {1} (\bibinfo {year} {2016})},\ \Eprint
  {https://arxiv.org/abs/1606.09551} {1606.09551} \BibitemShut {NoStop}%
\bibitem [{\citenamefont {Choudhary}\ \emph {et~al.}(2023)\citenamefont
  {Choudhary}, \citenamefont {Wines}, \citenamefont {Li}, \citenamefont
  {Garrity}, \citenamefont {Gupta}, \citenamefont {Romero}, \citenamefont
  {Krogel}, \citenamefont {Saritas}, \citenamefont {Fuhr}, \citenamefont
  {Ganesh} \emph {et~al.}}]{choudhary2023large}%
  \BibitemOpen
  \bibfield  {author} {\bibinfo {author} {\bibfnamefont {K.}~\bibnamefont
  {Choudhary}}, \bibinfo {author} {\bibfnamefont {D.}~\bibnamefont {Wines}},
  \bibinfo {author} {\bibfnamefont {K.}~\bibnamefont {Li}}, \bibinfo {author}
  {\bibfnamefont {K.~F.}\ \bibnamefont {Garrity}}, \bibinfo {author}
  {\bibfnamefont {V.}~\bibnamefont {Gupta}}, \bibinfo {author} {\bibfnamefont
  {A.~H.}\ \bibnamefont {Romero}}, \bibinfo {author} {\bibfnamefont {J.~T.}\
  \bibnamefont {Krogel}}, \bibinfo {author} {\bibfnamefont {K.}~\bibnamefont
  {Saritas}}, \bibinfo {author} {\bibfnamefont {A.}~\bibnamefont {Fuhr}},
  \bibinfo {author} {\bibfnamefont {P.}~\bibnamefont {Ganesh}}, \emph
  {et~al.},\ }\bibfield  {title} {\bibinfo {title} {Large scale benchmark of
  materials design methods},\ }\href@noop {} {\bibfield  {journal} {\bibinfo
  {journal} {arXiv preprint arXiv:2306.11688}\ } (\bibinfo {year}
  {2023})}\BibitemShut {NoStop}%
\bibitem [{\citenamefont {Smith}\ and\ \citenamefont
  {Topin}(2018)}]{smith2018superconvergence}%
  \BibitemOpen
  \bibfield  {author} {\bibinfo {author} {\bibfnamefont {L.~N.}\ \bibnamefont
  {Smith}}\ and\ \bibinfo {author} {\bibfnamefont {N.}~\bibnamefont {Topin}},\
  }\href@noop {} {\bibinfo {title} {Super-convergence: Very fast training of
  neural networks using large learning rates}} (\bibinfo {year} {2018}),\
  \Eprint {https://arxiv.org/abs/1708.07120} {arXiv:1708.07120 [cs.LG]}
  \BibitemShut {NoStop}%
\bibitem [{\citenamefont {Sanchez}\ \emph {et~al.}(1984)\citenamefont
  {Sanchez}, \citenamefont {Ducastelle},\ and\ \citenamefont
  {Gratias}}]{sanchez1984generalized}%
  \BibitemOpen
  \bibfield  {author} {\bibinfo {author} {\bibfnamefont {J.~M.}\ \bibnamefont
  {Sanchez}}, \bibinfo {author} {\bibfnamefont {F.}~\bibnamefont
  {Ducastelle}},\ and\ \bibinfo {author} {\bibfnamefont {D.}~\bibnamefont
  {Gratias}},\ }\bibfield  {title} {\bibinfo {title} {Generalized cluster
  description of multicomponent systems},\ }\href@noop {} {\bibfield  {journal}
  {\bibinfo  {journal} {Physica A: Statistical Mechanics and its Applications}\
  }\textbf {\bibinfo {volume} {128}},\ \bibinfo {pages} {334} (\bibinfo {year}
  {1984})}\BibitemShut {NoStop}%
\bibitem [{\citenamefont {Wang}\ \emph {et~al.}(2020)\citenamefont {Wang},
  \citenamefont {Li}, \citenamefont {Soisson},\ and\ \citenamefont
  {Becquart}}]{wang2020combining}%
  \BibitemOpen
  \bibfield  {author} {\bibinfo {author} {\bibfnamefont {Y.}~\bibnamefont
  {Wang}}, \bibinfo {author} {\bibfnamefont {K.}~\bibnamefont {Li}}, \bibinfo
  {author} {\bibfnamefont {F.}~\bibnamefont {Soisson}},\ and\ \bibinfo {author}
  {\bibfnamefont {C.~S.}\ \bibnamefont {Becquart}},\ }\bibfield  {title}
  {\bibinfo {title} {Combining dft and calphad for the development of
  on-lattice interaction models: The case of fe-ni system},\ }\href@noop {}
  {\bibfield  {journal} {\bibinfo  {journal} {Physical Review Materials}\
  }\textbf {\bibinfo {volume} {4}},\ \bibinfo {pages} {113801} (\bibinfo {year}
  {2020})}\BibitemShut {NoStop}%
\bibitem [{\citenamefont {Li}\ \emph {et~al.}(2022)\citenamefont {Li},
  \citenamefont {Fu}, \citenamefont {Nastar}, \citenamefont {Soisson},\ and\
  \citenamefont {Lavrentiev}}]{li2022magnetochemical}%
  \BibitemOpen
  \bibfield  {author} {\bibinfo {author} {\bibfnamefont {K.}~\bibnamefont
  {Li}}, \bibinfo {author} {\bibfnamefont {C.-C.}\ \bibnamefont {Fu}}, \bibinfo
  {author} {\bibfnamefont {M.}~\bibnamefont {Nastar}}, \bibinfo {author}
  {\bibfnamefont {F.}~\bibnamefont {Soisson}},\ and\ \bibinfo {author}
  {\bibfnamefont {M.~Y.}\ \bibnamefont {Lavrentiev}},\ }\bibfield  {title}
  {\bibinfo {title} {Magnetochemical effects on phase stability and vacancy
  formation in fcc fe-ni alloys},\ }\href@noop {} {\bibfield  {journal}
  {\bibinfo  {journal} {Physical Review B}\ }\textbf {\bibinfo {volume}
  {106}},\ \bibinfo {pages} {024106} (\bibinfo {year} {2022})}\BibitemShut
  {NoStop}%
\bibitem [{\citenamefont {Hattrick-Simpers}\ \emph {et~al.}(2023)\citenamefont
  {Hattrick-Simpers}, \citenamefont {Li}, \citenamefont {Greenwood},
  \citenamefont {Black}, \citenamefont {Witt}, \citenamefont {Kozdras},
  \citenamefont {Pang},\ and\ \citenamefont {Ozcan}}]{hattrick2023designing}%
  \BibitemOpen
  \bibfield  {author} {\bibinfo {author} {\bibfnamefont {J.}~\bibnamefont
  {Hattrick-Simpers}}, \bibinfo {author} {\bibfnamefont {K.}~\bibnamefont
  {Li}}, \bibinfo {author} {\bibfnamefont {M.}~\bibnamefont {Greenwood}},
  \bibinfo {author} {\bibfnamefont {R.}~\bibnamefont {Black}}, \bibinfo
  {author} {\bibfnamefont {J.}~\bibnamefont {Witt}}, \bibinfo {author}
  {\bibfnamefont {M.}~\bibnamefont {Kozdras}}, \bibinfo {author} {\bibfnamefont
  {X.}~\bibnamefont {Pang}},\ and\ \bibinfo {author} {\bibfnamefont
  {O.}~\bibnamefont {Ozcan}},\ }\bibfield  {title} {\bibinfo {title} {Designing
  durable, sustainable, high-performance materials for clean energy
  infrastructure},\ }\href@noop {} {\bibfield  {journal} {\bibinfo  {journal}
  {Cell Reports Physical Science}\ }\textbf {\bibinfo {volume} {4}} (\bibinfo
  {year} {2023})}\BibitemShut {NoStop}%
\bibitem [{\citenamefont {Cowley}(1950)}]{Cowley1950}%
  \BibitemOpen
  \bibfield  {author} {\bibinfo {author} {\bibfnamefont {J.~M.}\ \bibnamefont
  {Cowley}},\ }\bibfield  {title} {\bibinfo {title} {{An Approximate Theory of
  Order in Alloys}},\ }\href {https://doi.org/10.1103/PhysRev.77.669}
  {\bibfield  {journal} {\bibinfo  {journal} {Phys. Rev.}\ }\textbf {\bibinfo
  {volume} {77}},\ \bibinfo {pages} {669} (\bibinfo {year} {1950})}\BibitemShut
  {NoStop}%
\bibitem [{\citenamefont {Gong}\ \emph
  {et~al.}(2022{\natexlab{b}})\citenamefont {Gong}, \citenamefont {Xie},
  \citenamefont {Shao-Horn}, \citenamefont {Gomez-Bombarelli},\ and\
  \citenamefont {Grossman}}]{gong2022examining}%
  \BibitemOpen
  \bibfield  {author} {\bibinfo {author} {\bibfnamefont {S.}~\bibnamefont
  {Gong}}, \bibinfo {author} {\bibfnamefont {T.}~\bibnamefont {Xie}}, \bibinfo
  {author} {\bibfnamefont {Y.}~\bibnamefont {Shao-Horn}}, \bibinfo {author}
  {\bibfnamefont {R.}~\bibnamefont {Gomez-Bombarelli}},\ and\ \bibinfo {author}
  {\bibfnamefont {J.~C.}\ \bibnamefont {Grossman}},\ }\bibfield  {title}
  {\bibinfo {title} {Examining graph neural networks for crystal structures:
  limitations and opportunities for capturing periodicity},\ }\href@noop {}
  {\bibfield  {journal} {\bibinfo  {journal} {arXiv preprint arXiv:2208.05039}\
  } (\bibinfo {year} {2022}{\natexlab{b}})}\BibitemShut {NoStop}%
\bibitem [{\citenamefont {Dunn}\ \emph {et~al.}(2020)\citenamefont {Dunn},
  \citenamefont {Wang}, \citenamefont {Ganose}, \citenamefont {Dopp},\ and\
  \citenamefont {Jain}}]{Dunn2020}%
  \BibitemOpen
  \bibfield  {author} {\bibinfo {author} {\bibfnamefont {A.}~\bibnamefont
  {Dunn}}, \bibinfo {author} {\bibfnamefont {Q.}~\bibnamefont {Wang}}, \bibinfo
  {author} {\bibfnamefont {A.}~\bibnamefont {Ganose}}, \bibinfo {author}
  {\bibfnamefont {D.}~\bibnamefont {Dopp}},\ and\ \bibinfo {author}
  {\bibfnamefont {A.}~\bibnamefont {Jain}},\ }\bibfield  {title} {\bibinfo
  {title} {{Benchmarking materials property prediction methods: the Matbench
  test set and Automatminer reference algorithm}},\ }\href
  {https://doi.org/10.1038/s41524-020-00406-3} {\bibfield  {journal} {\bibinfo
  {journal} {npj Comput. Mater.}\ }\textbf {\bibinfo {volume} {6}},\ \bibinfo
  {pages} {1} (\bibinfo {year} {2020})}\BibitemShut {NoStop}%
\bibitem [{\citenamefont {Zhang}\ and\ \citenamefont
  {Gao}(2016)}]{zhang2016calphad}%
  \BibitemOpen
  \bibfield  {author} {\bibinfo {author} {\bibfnamefont {C.}~\bibnamefont
  {Zhang}}\ and\ \bibinfo {author} {\bibfnamefont {M.~C.}\ \bibnamefont
  {Gao}},\ }\bibfield  {title} {\bibinfo {title} {Calphad modeling of
  high-entropy alloys},\ }\href@noop {} {\bibfield  {journal} {\bibinfo
  {journal} {High-Entropy Alloys: Fundamentals and Applications}\ ,\ \bibinfo
  {pages} {399}} (\bibinfo {year} {2016})}\BibitemShut {NoStop}%
\bibitem [{\citenamefont {Chen}\ \emph {et~al.}(2018)\citenamefont {Chen},
  \citenamefont {Mao},\ and\ \citenamefont {Chen}}]{chen2018database}%
  \BibitemOpen
  \bibfield  {author} {\bibinfo {author} {\bibfnamefont {H.-L.}\ \bibnamefont
  {Chen}}, \bibinfo {author} {\bibfnamefont {H.}~\bibnamefont {Mao}},\ and\
  \bibinfo {author} {\bibfnamefont {Q.}~\bibnamefont {Chen}},\ }\bibfield
  {title} {\bibinfo {title} {Database development and calphad calculations for
  high entropy alloys: Challenges, strategies, and tips},\ }\href@noop {}
  {\bibfield  {journal} {\bibinfo  {journal} {Materials Chemistry and Physics}\
  }\textbf {\bibinfo {volume} {210}},\ \bibinfo {pages} {279} (\bibinfo {year}
  {2018})}\BibitemShut {NoStop}%
\bibitem [{\citenamefont {Merchant}\ \emph {et~al.}(2023)\citenamefont
  {Merchant}, \citenamefont {Batzner}, \citenamefont {Schoenholz},
  \citenamefont {Aykol}, \citenamefont {Cheon},\ and\ \citenamefont
  {Cubuk}}]{merchant2023scaling}%
  \BibitemOpen
  \bibfield  {author} {\bibinfo {author} {\bibfnamefont {A.}~\bibnamefont
  {Merchant}}, \bibinfo {author} {\bibfnamefont {S.}~\bibnamefont {Batzner}},
  \bibinfo {author} {\bibfnamefont {S.~S.}\ \bibnamefont {Schoenholz}},
  \bibinfo {author} {\bibfnamefont {M.}~\bibnamefont {Aykol}}, \bibinfo
  {author} {\bibfnamefont {G.}~\bibnamefont {Cheon}},\ and\ \bibinfo {author}
  {\bibfnamefont {E.~D.}\ \bibnamefont {Cubuk}},\ }\bibfield  {title} {\bibinfo
  {title} {Scaling deep learning for materials discovery},\ }\href@noop {}
  {\bibfield  {journal} {\bibinfo  {journal} {Nature}\ ,\ \bibinfo {pages} {1}}
  (\bibinfo {year} {2023})}\BibitemShut {NoStop}%
\end{thebibliography}%

\end{document}